\documentclass[9pt, a4paper, final, dvips, twoside]{extarticle}  

\usepackage[utf8]{inputenc}

\usepackage[headings,cm]{fullpage}

\usepackage[usenames]{color}

\usepackage{url}

\makeatletter
\def\url@leostyle{%
  \@ifundefined{selectfont}{\def\UrlFont{\sf}}{%
    \def\UrlFont{\rm}}\Url@do
}
\makeatother

\usepackage{times,tabularx,graphicx,pst-node,amssymb,amsmath,mdwlist,marvosym,fancyhdr}

\usepackage[afrikaans,british]{babel}

\usepackage{setspace}
\newcommand{\n}{'n~}
\newcommand{\ket}[1]{\ensuremath{|#1\rangle}}
\newcommand{\kol}[2]{\left(\!\begin{array}{rr}#1 \\ #2 \end{array}\!\right)}
\newcommand{\ckol}[2]{\left(\!\begin{array}{cc}#1 \\ #2 \end{array}\!\right)}
\newcommand{\matriks}[4]{\left(\!\begin{array}{rrrr}#1 \\ #2 \\ #3 \\ #4
\end{array}\!\right)}
\newcommand{\cmatriks}[4]{\left(\!\begin{array}{cccc}#1 \\ #2 \\ #3 \\ #4
\end{array}\!\right)}

\newcommand{\ken}{\ket{0}}
\newcommand{\kee}{\ket{1}}

\newcommand{\rem}[1]{}

\title{\flushleft \vskip -1cm
\Huge {\bf Kwantumberekening}\\
\large \bigskip
\rm \textsf{J Heidema}\\
\rm Departement Wiskundige Wetenskappe\\
\textsf{PH Potgieter\thanks{Outeur aan wie korrespondensie gerig kan
word.} ~en WL Fouché}
\\
\rm Departement Kwantitatiewe Bestuur, Universiteit van Suid-Afrika, Posbus 392, Unisa 0003\\
\rm E-pos: potgiph@unisa.ac.za
}

\author{}

\date{}

\begin{document}

\selectlanguage{afrikaans}

\hyphenation{ander-sins kwantum-ou-tomaat Cyber-netics teorie ver-plasing ek-sem-pla-rie-se be-skrywing waar-skyn-lik-heid kwantum-effekte waar-skyn-lik-heids-amplitudes uit-koms-toe-stande be-werk-stellig kwantum-berekening be-skou-ing me-tings-ingryping heel-alle}

\maketitle


\thispagestyle{fancy}

\noindent
{\bf \Large \textsc{Uittreksel}}\\
\emph{
Daar is geen gebrek aan dekking in die laaste paar jaar in wetenskaplike joernale van die onderwerp van kwantumberekening nie. Tereg ook, want dit is 'n nuwe idee met---tot sover---minstens één baie  belangrike praktiese toepassing (ontbinding in priemfaktore) sodra die tegnologie 
redelik omvattende berekenings kan hanteer. 
Die belangstelling in kwantumverwerking is aangevuur deur beide die besef dat die toenemende miniaturisering noodwendig rekenaarwetenskaplikes met kwantumverskynsels sal konfronteer, en deur  die begryp van die potensiaal van massief parallelle verwerking deur die gebruik maak van die  Hilbert-ruimtevoorstelling  van kwantumstelsels en die spesiale eienskappe van kwantumverstrengeling. Die onderwerp lê by die sameloop van fisika, rekenaarwetenskap en wiskunde en hierdie bydrae verken die idees van dié drie gebiede wat die belangrikste rol in kwantumberekening speel en besin oor die definisies van en die waarskynlike gebruik vir kwantumverwerking asook sy  verhouding tot die klassieke berekeningskompleksiteit. Ten slotte word 'n uitgebreide geannoteerde bibliografie verskaf as gids vir die geïnteresseerde leser tot die literatuur in dié onderwerp.
}

\bigskip

\selectlanguage{british}\noindent
{\Large\bf\textsc{Abstract}}\\
{\bf\emph{Quantum computing}}\\
\emph{
There has been no lack of coverage in the past few years in scientific journals of the topic of quantum computation. Rightly so, as this is a novel idea with---so far---at least one very important practical application (prime factorisation) as soon as the technology can accommodate reasonably  sized computations. Interest in quantum computing has been sparked by both the realization that increasing miniaturisation will eventually bring computer scientists face-to-face with quantum effects, and by the grasping of the potential for massive parallelism inherent in the Hilbert space representation of quantum systems and in the special effects of quantum entanglement. 
The topic lies at the confluence of physics, computer science and mathematics and this contribution reviews the ideas from these three fields that play the most important role in quantum computation and reflects on the definitions of and prospective use for quantum computing as well as its relation to classical computational complexity. Finally, an extended annotated bibliography is provided as a guide for the interested reader to the literature in the subject.
}

\selectlanguage{afrikaans}


\section{\uppercase{Inleiding}}

Let 'n mens op publikasies in die kwantumfisika, dan val dit op dat daar nou
'n vloedgolf van artikels verskyn oor kwantumberekening en -rekenaars, asook
oor verwante sake soos kwantuminligting, \mbox{-kommunikasie}, -kriptografie en
\mbox{-verplasing} ("teleportasie"). In hierdie artikel word die grondbeginsels van
kwantumberekening uiteengesit en eenvoudig ge\"{\i}llustreer.
Om alles in besonderhede te volg, is soveel kennis van die line\^ere algebra
(vektore en matrikse) nodig as wat gewoonlik in twee jaar se wiskunde op
universiteit geleer word.

Kwantumberekening behels méér as net die feit dat komponente van rekenaars so klein gemaak word dat kwantumeffekte 'n rol begin speel. Dit benut op 'n wesenlike wyse dié kenmerkende eienskap van 'n saamgestelde kwantumstelsel, naamlik die moontlikheid van \emph{verstrengelde} toestande van die geheel, wat \emph{gelyktydig} 'n groot aantal verskillende konfigurasies van toestande van komponente interfererend laat ontwikkel---en 'n uitkoms kan laat \emph{bereken}, 
as die stelsel reg ontwerp en gebruik word. Daar is, teoreties gesproke, groot belofte in kwantumberekening, maar ook, prakties gesproke, groot probleme vir die kwantumingenieur.

\section{\uppercase{Klassieke Berekening}}

Wat is, in die algemeen, 'n \textit{berekening}? Ons het op skool sommige
soorte van berekening bemeester, soos die algoritmes vir vermenigvuldiging en
langdeling van getalle. Allerlei hulpmiddels vergemaklik berekening: pen en
papier, 'n abakus, 'n rekenaar. Maar elke berekening, hoe dit ook al uitgevoer
word, is 'n proses waarin 'n eindige string ("woord", rytjie) simbole omskep
word in 'n soortgelyke string. Nou praat ons van "klassieke" berekening, wat
nie die kwantumteorie betrek nie. 'n String is dan 'n eindige ry van simbole
uit 'n eindige alfabet, terwyl mens, om oop ruimte te skep vir die bewerking van
die string, mag aanneem dat enige aantal blanko simbole (m.a.w. "niks", oop
plekke) voor en agter aan die string gelas mag word:%

\[%
\begin{tabular}
[c]{c|c|c|c|c|c|c|c|c|c}\hline
$\cdots$ & \ \ \ \  & \ \ \ \  & $\ 1\ $ & $\ 1\ $ & $0\ $ & $\ 1\ $ &
\ \ \ \  & \ \ \ \  & $\cdots$\\\hline
\end{tabular}
\]

\medskip
\noindent 
Hierdie string kan ons kortweg skryf as $(1;1;0;1),$ of sommer $1101$.

Die string waarmee 'n berekening begin, noem ons soms die \textit{invoer}, of
gegewene, premisse, probleem, data, ens., afhangende van die aard van die
berekening wat daarop toegepas moet word. Die string waarmee die
berekeningsproses eindig, noem ons die \textit{afvoer}, resultaat, konklusie,
antwoord, beslissing, ens., terwyl tussenstappe in die berekening
\textit{deurvoere} lewer. Die berekeningsproses self, wat die invoer (gewoonlik
met tussenstappe) transformeer tot die afvoer, noem ons soms 'n
\textit{prosedure}, algoritme, of program, waarby onderskei moet word tussen
'n \textit{beskrywing} of spesifikasie van die prosedure (gewoonlik vir
verskillende moontlike invoere) en 'n \textit{uitvoering} van die prosedure
(vir een besondere invoer).

Die uitvoering van 'n prosedure (program, "sagteware", selfs
"Turing-\textit{masjien}", op 'n spesifieke invoer) in die fisiese werklikheid
verg 'n fisiese stelsel, 'n \textit{rekenaar}, mens of masjien, asook die
volgende verbande: die invoer word gerepresenteer deur 'n fisiese toestand van
('n deel van) die rekenaar; ooreenkomstig die prosedure verander die toestand
van die rekenaar, gewoonlik stap vir stap, om die opeenvolgende deurvoere te
representeer soos hulle deur die "stroombaan" van die rekenaar vloei; en die
rekenaar bereik 'n eindtoestand wat (met 'n deel daarvan) die afvoer fisies
representeer. Die program self word gewoonlik ook fisies in die rekenaar
gerepresenteer as deel van die rekenaar se toestand en lei die hele prosedure
van begintoestand tot eindtoestand. 'n "Meerdoelige" of "universele" rekenaar
kan verskillende programme uitvoer, terwyl 'n "eendoelige" rekenaar (dalk 'n
"module" binne-in 'n universele rekenaar) een program uitvoer, wat in sy
stroombaan vasgel\^{e} ("hardwired") is.

Ons gee twee eenvoudige voorbeelde van berekening. 'n Voorbeeld uit die
element\^{e}re rekenkunde:%

\[
137 \times291 = 39867.
\]

Die invoer hier is die string $137;291$. Die simbole $\times$ en $=$ dui op die
transformasieprosedure "vermenigvuldiging". Die afvoer is die string $39867$.

'n Voorbeeld uit die proposisielogika:%
\[%
\begin{tabular}
[c]{l}%
$p$\\
$\underline{p\rightarrow q}$\\
$\therefore\ q$%
\end{tabular}
\]

\medskip

Die invoer is die string $p;p\rightarrow q$, genoem die premisse; die afvoer is
$q$, genoem die konklusie; en die horisontale streep en die simbool
$\therefore$ dui op die transformasieprosedure "logiese afleiding", wat s\^{e}
dat as ons die bewering "$p$" weet of aanneem, asook "as $p$, dan $q$", dan
mag ons die gevolgtrekking "$q$" maak. Ons het hier voorbeelde van
berekeningsprosesse wat op verskillende invoere van toepassing is: as ons
weet hoe om te vermenigvuldig (waarvoor ons die tussenstappe nie hierbo
uitstippel nie!), dan kan ons enige twee getalle vermenigvuldig; en as ons
bostaande logiese deduksiere\"{e}l (wat logici \textit{modus ponens} noem,
sonder tussenstappe) ken, dan kan ons dit op enige twee bewerings $p$ en $q$
toepas.

Ons merk op dat die alfabet van simbole beperk mag word tot $\{0;1\}$ (of
$\{0;1;b\}$, as die blanko simbool 'n naampie verdien). Enige eindige string
oor enige eindige alfabet kan gekodeer word as 'n eindige string van nulle en
ene --- soos dit trouens fisies gerepresenteer word in ons gewone rekenaars.

Die wiskundige Alan Turing het in 1936 'n skerpsinnige konseptuele ontleding gemaak
van berekening, d.w.s. die transformasieprosedure van 'n invoerstring na 'n
afvoerstring, wat aanneemlik maak dat enige berekening op die simboliese vlak
uitgevoer kan word deur 'n eenvoudige soort abstrakte "rekenaar", bekend as 'n
\textit{Turing-masjien} (kortweg TM), wat opeenvolgende element\^{e}re stappe
in die transformasie van die string uitvoer en dan halt roep as die
afvoerstring bereik is. Intu\"{\i}tief kan ons aan 'n TM en sy funksionering
soos volg dink: Die eindige (invoer-, deurvoer-, en afvoer-) stringe van nulle
en ene kom voor op 'n line\^{e}re band, potensieel oneindig lank na links en
na regs, verdeel in selle (hokkies) wat elkeen een simbool bevat of blanko is.
Die TM het 'n eindige aantal toestande of interne konfigurasies, en 'n "kop",
wat tydens elke stap van die berekening na net een van die selle op die band
"kyk" en daardie sel se simbool $(0,1\ \text{of}\ b)$ lees. Die TM het ook 'n
eindige lys instruksies of oorgangsre\"els, 'n program, wat vir elke moontlike 
toestand $q_{i}$
en vir elk van die simbole wat die kop kan sien, s\^{e} $S_{j}$, \'{o}f geen
instruksie bevat nie---in welke geval die TM dadelik halt wanneer dit in toestand 
$q_i$ simbool $S_j$ sien---\'{o}f 'n
instruksie bevat van presies een van die volgende drie vorme:
\begin{tabbing}
\quad \= (1) \quad \= $q_{i}S_{j}S_{k}q_{\ell}$ \= : \= vervang simbool\ $S_{j}\ $in die sel waarna jy kyk met simbool $S_{k}$ en gaan dan oor in toestand $q_{\ell},$ terwyl\\
\> \> \> \>  jy na dieselfde sel bly kyk; \\
\> (2) \> $q_{i}S_{j}Rq_{\ell}$ \> : \> los simbool $S_{j}\ $waar dit is, beweeg die kop een sel na regs
en gaan oor in toestand $q_{\ell};$\\
\> (3) \> $q_{i}S_{j}Lq_{\ell}$ \> : \> soortgelyk, maar nou een sel na links.
\end{tabbing}
So bv. beteken die instruksie $q_{1}b1q_{2}$: as jy in toestand $q_{1}$ is en
jou kop kyk na 'n blanko sel, skryf 'n $1$ in daardie sel, hou aan om na
daardie sel te kyk, en gaan oor in toestand $q_{2}$.

\medskip

'n \textit{Berekening }van die TM begin, bv., met die kop oor die eerste
nie-blanko sel van links in die \textit{invoerstring} en die masjien in
toestand $q_{0}$. Hy sien in daardie sel bv. 'n $0$. In sy program vind hy die
opdrag $q_{0}0\,1q_{2}$. Hy verander die $0$ na $1$ en gaan in toestand
$q_{2}$. Nou is hy in toestand $q_{2}$, sien 'n $1$, en moet die volgende stap
in die berekening uitvoer. Hy soek 'n opdrag $q_{2}1\cdot\cdot$\ in sy
program. As hy dit vind, voer hy dit uit; ensovoorts. As die TM in toestand
$q_{i}$ die simbool $S_{j}$ in die betrokke sel sien en geen opdrag
$q_{i}S_{j}\cdot\cdot\ $in sy program vind nie, dan halt hy. Die string op die
band as die TM halt, is die \textit{afvoerstring}, waarna die berekening eindig vir
die invoerstring waarmee begin is. ('n "Swak" program kan natuurlik 'n TM in 'n
oneindige lus laat beland wat nooit halt nie!)

\medskip

\noindent Om saam te vat: 'n berekening deur 'n TM, as transformasie van 'n
invoerstring tot 'n afvoerstring, vind in 'n eindige opeenvolging van
tussenstappe plaas, elk waarvan net een simbool betrek en bepaal word deur een
instruksie in die TM se program, waarna in die volgende stap dieselfde sel of
die een net links of net regs daarvan aan die beurt kom. Daar bestaan goeie
redes om te aanvaar dat enige prosedure wat ons 'n (klassieke) berekening sou
noem, deur so 'n TM-berekening gesimuleer kan word. 'n Mens kan selfs op 'n
konstruktiewe wyse bewys dat daar 'n \textit{universele} TM bestaan, een
Turing-masjien, s\^{e} $T$, wat die berekeningsgedrag van \textit{enige
}Turing-masjien $X$ op die volgende wyse kan simuleer: 'n invoerstring vir $T$
bestaan uit twee deelstringe, nl. die invoerstring $Y$ waarop die berekening
(soos deur $X$) gedoen moet word, en 'n deelstring wat $X$ beskryf. $T$ doen
dan met $Y$ wat (die "program") $X$ daarmee sou doen en produseer di\'{e}
afvoerstring wat $X$ sou lewer. Ons gewone tafelrekenaar is dan vir
berekeningsdoeleindes ekwivalent aan 'n universele TM --- behalwe dat
eersgenoemde net 'n eindige "band" (geheue en werkruimte) het, teenoor die TM
se potensieel onbeperkte aantal selle op sy band.

Bostaande uitleg van \textit{klassieke berekening as berekening deur 'n
(universele) Turing-masjien} is eenvoudig en fundamenteel en spreek sterk tot
ons intu{\"{\i}}sie oor wat berekening is. Daar bestaan ook ander,
ekwivalente, formele rekonstruksies van die informele begrip berekenbaarheid.
Op sommige daarvan, soos algemene rekursiwiteit of $\lambda$%
-definieerbaarheid, wil ons nie hier ingaan nie. Maar ons wil 'n ander, ook
ekwivalente, wiskundige beskrywing van berekening gee, wat aan die een kant
sterk aansluit by Turing se benadering (manipulasie van stringe simbole),
maar, aan die ander kant, klassieke berekening beskryf met presies di\'{e}
soort wiskundige parafernalia wat noodsaaklik is vir die beskrywing van
kwantumberekening.

\section{\uppercase{Klassieke berekening met vektore}}

\noindent
'n Berekening kan ook ontleed word as bestaande uit meer \textit{algemene}
stappe as di\'{e} van 'n TM, stappe waarby die \textit{hele string} in elke
stap getransformeer word en nie net hoogstens een simbool per stap nie. Aan
die ander kant gaan ons ontleding nou meer \textit{beperkend }raak as di\'{e}
met 'n TM, want die stringe gaan as vektore in 'n vektorruimte gesien word wat
slegs \textit{line\^{e}re} transformasies ondergaan. Dit beteken dat in elke
berekeningstap die nuwe string op 'n heel spesifieke algebra\"{\i}ese wyse uit
die oue gekonstrueer word --- en nie deur sommer enige eensellige verandering,
soos deur 'n TM toegelaat nie. Om meer presies te wees: daar kan bewys word
dat enige formele (bv. TM-) berekening soos volg logies gerekonstrueer kan
word:

\begin{itemize}
\item die invoer-, deurvoer- en afvoerstringe is vektore in 'n sekere
vektorruimte;

\item die berekening, invoer $\mapsto$ afvoer, word uitgevoer deur die
invoervektore deur 'n logiese "stroombaan" (met seri\"{e}le en parallelle
skakeling) te stuur;

\item elke node in die stroombaan (gesien as 'n gerigte grafiek) is 'n
omkeerbare logiese hek;

\item 'n omkeerbare logiese hek moet, in hierdie konteks, 'n nie-singuliere
line\^{e}re transformasie wees, voorgestel deur 'n inverteerbare matriks.
\end{itemize}

Dit alles sal ons nou in meer besonderhede verduidelik.

\medskip

\noindent Aan die versameling van alle eindige stringe (van enige eindige
lengte) van nulle en ene, kan ons dink as 'n vektorruimte, formeel die
tensorsom van alle eindigdimensionele vektorruimtes oor die liggaam van
skalare $\mathbb{Z}_{2}=\{0;1\}$. $\mathbb{Z}_{2}$ het die volgende
rekenkunde:%

\begin{align*}
0+0  & =0\\
0+1  & =1\ =\ 1+0\\
1+1  & =0\\
-a  & =a\\
0\times a  & =0\ =\ a\times0\\
1\times a  & =a\ =\ a\times1
\end{align*}

\medskip

\noindent Die twee simbole $0$ en $1$ staan bekend as \textit{bisse }(vir
"bin\^{e}re syfers", in Engels "bits" vir "binary digits").

Om die verduideliking eenvoudig, maar darem nie heeltemal triviaal nie, te
hou, kies ons hoofsaaklik voorbeelde uit en definieer begrippe vir die
eksemplariese deelruimte $V_{2}=\{(0;0);(0;1);(1;0);(1;1)\}$ van die vier
tweedimensionele vektore (stringe van bisse) oor $\mathbb{Z}_{2}$;
$V_{2}=\{(x;y)\,|\,x,y\in\mathbb{Z}_{2}\}$.

'n Algemene \textit{logiese hek} of \textit{Boole-funksie }is enige funksie
$h:V_{m}\rightarrow V_{n}$. Daar is $4^{4}=256$ logiese hekke $h:V_{2}%
\rightarrow V_{2}$ (vir stringe van lengte twee), want elk van die vier
vektore kan op enigeen van hulle afbeeld. 'n\textit{\ Omkeerbare} logiese hek
is een waarvan die inverse ook 'n logiese hek is, d.w.s. 'n bijeksie
(injektiewe en surjektiewe funksie) $h:V_{2}\rightarrow V_{2}$. Van die $256$
hekke is net $4!=4\times3\times2\times1=24$ omkeerbaar, (want geen twee
verskillende stringe mag op dieselfde een afbeeld nie, sodat die invoer
eenduidig vanuit die afvoer herwin kan word). 'n Logiese hek wat nie
omkeerbaar is nie, verloor inligting wanneer 'n string deur hom gaan, omdat sy
invoer nie (altyd, volledig) vanuit sy afvoer gerekonstrueer kan word nie. 'n
Fisiese stelsel wat 'n nie-omkeerbare logiese hek se invoer-afvoer-gedrag
realiseer, bring ook noodwendig volgens sommige kenners hitte voort, d.w.s. verkwiste energie wat 'n
oorlas word.
In die 1950's, byvoorbeeld, het John von Neumann geskryf dat elke logiese bewerking deur 'n rekenaar by
konstante temperatuur 'n vaste minimum-hoeveelheid energie moet verstrooi. Dat dit egter nié so is nie,
is al in 1961 deur Rolf Landauer uitgewys. Volgens Landauer was dit nie berekening op sigself nie, maar die uitwis
van inligting wat die hitte sou genereer. Charles Bennett toon in 1973 aan dat alle Turing-masjienberekening deur
\emph{omkeerbare} Turing-masjiene gedoen kan word, sonder hitteverstrooiing. Kwantumberekening is uit die aard van die saak (sien later) omkeerbaar en vermy in beginsel die hitteverstrooiing wat met gewone berekening gepaardgaan.

Ons kan die stringetjie simbole $xy$ voorstel deur die \emph{ryvektor} $(x;y)$,
maar wanneer ons die vektore van \emph{links} met matrikse vermenigvuldig moet ons
die, getransponeerde, \emph{kolomvektor} gebruik:
$$(x;y)^T = \kol{x}{y}.$$
'n \textit{Line\^{e}re} logiese hek is 'n line\^{e}re transformasie
$a:V_{2}\rightarrow V_{2},\ (x;y)^T \mapsto(x^{\prime};y^{\prime})^T$, wat
voorgestel kan word as vermenigvuldiging van die kolomvektor van links met 'n
matriks $A$; $A(x;y)^T=(x^{\prime};y^{\prime})^T$, of%
\[
A\kol{x}{y}  =  \left(
\begin{array}
[c]{rr}%
a_{11} & a_{12}\\
a_{21} & a_{22}%
\end{array}
\right)
\kol{x}{y}
=\left(
\begin{array}
[c]{c}%
a_{11}x+a_{12}y\\
a_{21}x+a_{22}y
\end{array}
\right)  =\kol{x^{\prime}}{y^{\prime}}  .
\]

Van die $256$ logiese hekke is $16$ line\^{e}r, want daar is $16$ matrikse met
vier inskrywings $a_{ij}$ uit $\mathbb{Z}_{2}$. Van hierdie $16$ line\^{e}re
hekke is $10$ singulier, ofte wel nie omkeerbaar nie. Dit gebeur as
$\det(A)=0$, wat die geval is as $A$ vier of drie nulle het, of twee nulle in
dieselfde ry of kolom, of geen nulle nie. Die $6$ omkeerbare line\^{e}re
logiese hekke het nie-singuliere (inverteerbare) matrikse met determinant $1$.
Ons noem di\'{e} $6$ matrikse $I,J,K,L,M,N$ en vertel hoe hulle 'n string
$(x;y)^T$ (enigeen van die vier) verwerk tot $(x^{\prime};y^{\prime})^T$:%

\begin{align*}
I\kol{x}{y}   & =  \left(
\begin{array}
[c]{rr}%
1 & 0\\
0 & 1
\end{array}
\right) \kol{x}{y} =\kol{x}{y}  ;\\
& \\
J\kol{x}{y}   & =  \left(
\begin{array}
[c]{rr}%
0 & 1\\
1 & 0
\end{array}
\right) \kol{x}{y} =\kol{y}{x} ;\\
& \\
K\kol{x}{y}  & =  \left(
\begin{array}
[c]{rr}%
1 & 1\\
1 & 0
\end{array}
\right) \kol{x}{y} =\ckol{x+y}{x}  ;\\
& 
\end{align*}
\begin{align*}
L\kol{x}{y}   & = \left(
\begin{array}
[c]{rr}%
0 & 1\\
1 & 1
\end{array}
\right) \kol{x}{y}  =\ckol{y}{x+y}  ;\\
& \\
M\kol{x}{y} & = \left(
\begin{array}
[c]{rr}%
1 & 1\\
0 & 1
\end{array}
\right) \kol{x}{y}  =\ckol{x+y}{y}  ;\\
& \\
N\kol{x}{y}   & =\left(
\begin{array}
[c]{rr}%
1 & 0\\
1 & 1
\end{array}
\right) \kol{x}{y}   =\ckol{x}{x+y}.
\end{align*}
Hierdie ses matrikse vorm 'n groep onder matriksvermenigvuldiging, wat weer
die seri\"{e}le skakeling van hekke voorstel:%

\[
(BA)\kol{x}{y} = B\left[ A\kol{x}{y}\right] ;
\]
eers word $(x;y)^T$ deur hek $A$ gejaag en die resultaat daarna deur
hek $B$. Maar elke produk van enige aantal van die ses hekke is weer een van
hulle, bv. $K(LM)=KJ=M$. Elke direk opeenvolgende skakeling van enige aantal
in enige seri\"{e}le orde is dus ekwivalent met 'n enkele hek. Verder is
$I,J,M$ en $N$ elkeen sy eie inverse, terwyl $K$ en $L$ mekaar se inverses is.
Verskeie ondergroepe met twee of drie (die delers van ses) elemente bestaan.

Die hek $N:(x;y)^T\mapsto(x;x+y)^T$ heet die "beheerde-nie-hek" (\emph{BNIE}, in Engels
\emph{CNOT} of "controlled not-gate") en kan skematies in 'n stroombaan soos volg
voorgestel word:%

\[%
\begin{tabular}
[c]{c|c|l}\cline{2-2}%
\begin{tabular}
[c]{c}%
$x\rightarrow$\\
$y\rightarrow$%
\end{tabular}
& $\ N\ $ &
\begin{tabular}
[c]{l}%
$\rightarrow x$\\
$\rightarrow x+y$%
\end{tabular}
\\\cline{2-2}%
\end{tabular}
\]
Omdat $0$ en $1$ in die logika "onwaar" en "waar" voorstel, kan ons die hek
$N$ ook definieer deur die volgende "waarheidstabel":%

\[%
\begin{tabular}
[c]{|c|c|c|c|}\hline
$\ x\ $ & $\ y\ $ & $\ x^{\prime}$ & $\ y^{\prime}$\\\hline
$0$ & $0$ & $0$ & $0$\\
$0$ & $1$ & $0$ & $1$\\
$1$ & $0$ & $1$ & $1$\\
$1$ & $1$ & $1$ & $0$\\\hline
\end{tabular}
\]
Let op dat
\begin{itemize}
\item $x=x^{\prime}$ onveranderd deur die hek gaan;
\item $y^{\prime}=x+y$, waar $+$ optelling modulo $2$ in $\mathbb{Z}_{2} $
is, maar ook as die logiese bewerking "eksklusiewe disjunksie" (in Engels \emph{XOR})
gesien kan word, m.a.w. "$x$ of $y$, maar nie beide nie";
\item $x=0\Rightarrow y=y^{\prime}$ en $x=1\Rightarrow y\not =y^{\prime} $,
sodat $x$ se waarde dus "beheer" of die "teiken" $y$ dieselfde moet bly of in nie-$y$ moet
verander.
\end{itemize}
'n Ander eienskap wat 'n omkeerbare line\^{e}re logiese hek (nie-singuliere
matriks) kan h\^{e}, is \textit{unitariteit}. By kwantumberekening word
hierdie begrip essensieel. Daar word dit wel anders gedefinieer (omdat ons
daar met vektorruimtes oor die komplekse getalle $\mathbb{C}$ sal werk), maar
in ons eenvoudige klassieke konteks is matriks $A$ \textit{unit\^{e}r }as
$A^{-1}=A^{T}$, m.a.w. $A$ se inverse is sy getransponeerde, daardie matriks
waarvan, vir elke $i$, sy $i$-de kolom die $i$-de ry van $A$ is. Onder ons ses
betrokke matrikse is net twee unit\^{e}r, nl. $I$ en $J.$
\n Unit\^ere matriks behou die norm (lengte) van elke vektor waarop hy inwerk. 

'n Logiese stroombaan hoef hekke nie net serieel te skakel nie, maar die baan
kan splits (vertak, uitwaaier) in parallelle takke, waarby dieselfde string
parallel na twee of meer hekke gestuur word. Dit kan sommer gedoen word deur
die regte konfigurasie van "drade" vir die stroombaan. Maar as mens dit tog
wiskundig wil voorstel, is vermenigvuldiging met 'n ander soort matriks
gepas, bv.%

$$
\kol{I}{I} \kol{x}{y}  =
\matriks{1&0}{0&1}{1&0}{0&1}
\kol{x}{y}  =\matriks{x}{y}{x}{y} .
$$

\[%
\begin{tabular}
[c]{c|c|l}\cline{2-2}%
$%
\begin{array}
[c]{r}%
x\rightarrow\\
y\rightarrow
\end{array}
$ & $\kol{I}{I}  $ & $%
\begin{array}
[c]{r}%
\rightarrow x\\
\rightarrow y\\
\\
\rightarrow x\\
\rightarrow y
\end{array}
$\\\cline{2-2}%
\end{tabular}
\]
Om die twee (of meer) parallelle drade weer saam te voeg, voer ons hulle
eenvoudig in dieselfde hek in (waarby ons dan natuurlik ook weer na groter
matrikse as $2\times2$ moet gaan om daardie hek voor te stel).

Nou moet ons dadelik daarop wys dat die klassieke verdubbeling van 'n string bisse om 'n stroombaan in parallelle takke te splits, in kwantumberekening nie moontlik is nie. Dit volg uit die sogenaamde Kloonverbod-stelling wat sê dat dit fisies onmoontlik is om 'n bestaande onbekende kwantumbis te kopieer (\ref{FK}). Gevolglik moet mens in kwantumberekening van die begin af 'n invoerregister skep wat lank genoeg is om alle nodige kopieë van bisse in een vektor saam te dra deur die seriële proses (onvertakte stroombaan) van kwantumberekening.

As ons $m$ en $n$ al hul moontlike waardes laat aanneem, dan is daar oneindig veel logiese hekke $h:V_m\rightarrow V_n$. In die logika is dit lankal bekend dat 'n mens enige logiese hek kan verkry deur samestellings van 'n paar eenvoudige hekke, bv. van \emph{nie} (ontkenning), $\neg:V_1\rightarrow V_1$, en \emph{en} (konjunksie), $\wedge:V_2\rightarrow V_1$, met waarheidstabelle

\begin{center}
\begin{tabular}{|c|c|}
\hline $x$ & $\neg x$ \\
\hline 0 & 1 \\
\hline 1 & 0 \\
\hline
\end{tabular}
\quad en \quad
\begin{tabular}{|c|c|c|}
\hline $x$ & $y$ & $x\wedge y$ \\
\hline 0 & 0 & 0 \\
\hline 0 & 1 & 0 \\
\hline 1 & 0 & 0 \\
\hline 1 & 1 & 1 \\
\hline
\end{tabular}~.
\end{center}
Die hek \emph{of} (inklusiewe disjunksie), $\vee:V_2 \rightarrow V_1$, het bv. presies dieselfde waarheidstabel as 'n hek opgebou uit $\neg$ en $\wedge$: as hekke gesien, het ons $x\vee y = \neg(\neg x \wedge \neg y)$. 'n Mens kan selfs bewys dat daar twee hekke bestaan, $\mid,\downarrow:V_2\rightarrow V_1$, respektiewelik bekend as die \emph{Sheffer-haal} of \emph{NEN} (\emph{Sheffer stroke} of \emph{NAND} in Engels) en die \emph{dolk} of \textit{NOF}  (\emph{dagger} of \textit{NOR}), met waarheidstabelle
\begin{center}
\begin{tabular}{|c|c|c|}
\hline $x$ & $y$ & $x\mid y$ \\
\hline 0 & 0 & 1 \\
\hline 0 & 1 & 1 \\
\hline 1 & 0 & 1 \\
\hline 1 & 1 & 0 \\
\hline
\end{tabular}
\quad en \quad
\begin{tabular}{|c|c|c|}
\hline $x$ & $y$ & $x\downarrow y$ \\
\hline 0 & 0 & 1 \\
\hline 0 & 1 & 0 \\
\hline 1 & 0 & 0 \\
\hline 1 & 1 & 0 \\
\hline
\end{tabular}~,
\end{center}
wat elkeen op sigself enige hek kan voortbring; bv. $\neg x = x\mid x = x\downarrow x$; $x\wedge y = (x\mid y)\mid(x\mid y) = (x\downarrow x)\downarrow(y \downarrow y)$. (Soos so dikwels met vernoemings gebeur, was professor H.M. Sheffer nié die eerste persoon om die haal te gebruik nie, maar wel die filosoof C.S. Peirce.) 

Maar natuurlik is nie een van $\wedge$, $\mid$ of $\downarrow$ 'n omkeerbare logiese hek nie. Die vraag ontstaan dus of daar 'n eindige versameling van \emph{omkeerbare} logiese hekke, miskien selfs 'n enkele een, bestaan wat alle omkeerbare logiese hekke deur samestelling kan voortbring. Die antwoord is \emph{ja}, daar is omkeerbare hekke wat op hul eie enige omkeerbare hek kan lewer, maar geen tweebis-hek kan dit doen nie. Mens moet na ten minste drie bisse gaan. So is daar die \emph{Fredkin-hek} en die beheerde-beheerde-nie-hek \emph{BBNIE} (sien later oor sy lineariteit en unitariteit), ook bekend as die \emph{Toffoli-hek} $T:V_3\rightarrow V_3$  (1981), wat só werk:
\[%
\begin{tabular}
[c]{c|c|l}\cline{2-2}%
$%
\begin{array}
[c]{r}%
x\rightarrow\\
y\rightarrow\\
z\rightarrow
\end{array}
$ & {\large T} & $%
\begin{array}
[c]{l}%
\rightarrow x\\
\rightarrow y\\
\rightarrow z + xy
\end{array}
$\\\cline{2-2}%
\end{tabular}
\]

Let ook op dat as ons, in die invoerdrietal van die Toffoli-hek, $z$ vaspen op $z=1$, dan is $z'=1+xy=x|y=x\mbox{~NEN~}y$. Dit beteken dat enige klassieke logiese hek, selfs die onomkeerbares, nageboots kan word deur 'n stroombaan met slegs klassieke Toffoli-hekke waarvan op die $z$-draad altyd net 1 ingevoer word.

Wanneer ons deur die bril van die kwantumfisika na berekening kyk as 'n
prosedure wat deur 'n fisiese stelsel uitgevoer moet word, dan besef ons dat
nie net stringe van bisse nie, maar selfs die enkel bisse, $0$ en $1,$ beide
deur \textit{nie-nul }vektore voorgestel moet word. Dit sal dus nie deug om
hulle in kwantumberekening as die twee vektore in $V_{1}\ $te sien nie. Die
rede hiervoor is die volgende: 'n bis, klassiek wiskundig voorgestel deur $0$
of $1,$ word klassiek fisies gerealiseer deur ('n deel van) 'n fisiese stelsel
wat ondubbelsinnig in een van twee duidelik onderskeibare makrotoestande
verkeer:\ die simbool wat ek op die papier lees of skryf is \'{o}f $0$, \'{o}f
$1$; die lig brand \'{o}f nie, \'{o}f wel; die skakelaar is \'{o}f af, \'{o}f
aan; daar is \'{o}f nie, \'{o}f wel 'n ponsgaatjie; die transistor gelei
\'{o}f nie, \'{o}f wel; ens. Ons het dus 'n bi-stabiele komponent wat in een
van twee maklik lees- of meetbare \textit{toestande }is.

Nou is dit so dat in die kwantumteorie elke fisiese stelsel wiskundig
voorgestel word deur 'n sekere paslike vektorruimte, en dat elke toestand van
die stelsel, of van 'n komponent daarvan, met 'n \textit{nie-nul }vektor
geassosieer word. Die nulvektor verteenwoordig "onmoontlikheid", iets wat nie
fisies of wiskundig met die stelsel kan gebeur nie. Wat meer is, die twee
fisiese toestande wat $0$ en $1,$ "nee" en "ja", "onwaar" en "waar" realiseer,
word wiskundig voorgestel deur twee ortogonale vektore "loodreg" op mekaar.
Om dus die twee klassieke bisse wiskundig voor te stel op die wyse wat netnou
in die kwantumberekening nodig word, gaan ons hulle

\begin{itemize}
\item nuwe name gee, naamlik $\left\vert 0\right\rangle \ $en\ $\left\vert
1\right\rangle ,\ $om aan te dui dat hulle nou \textit{vektore }is (met Dirac
se \textit{ket}-notasie); en

\item soos volg definieer as twee \textit{ortogonale }vektore in $V_{2}\ $van
\emph{norm} ("lengte") een, 'n \emph{ortonormale basis}, as u wil:%
\[
\left\vert 0\right\rangle :=\kol{1}{0} \ \ \text{en\ \ }\left\vert
1\right\rangle :=\kol{0}{1} .
\]

\end{itemize}%

\[
\includegraphics{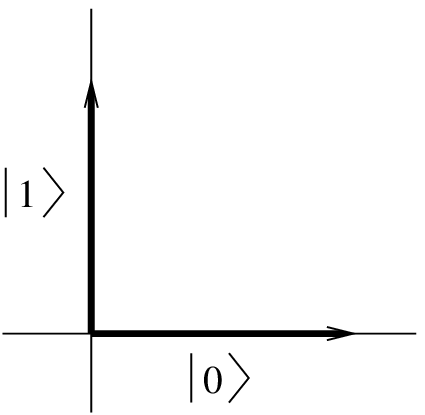}
\]

Ons moet hierby egter net onthou dat optelling van bisse soos tevore plaasvind
en nie ooreenkom met komponentsgewyse optelling in $V_{2}\ $nie, bv.%
\begin{align*}
\left\vert 0\right\rangle +\left\vert 1\right\rangle  & =\left\vert
1\right\rangle ,\ \ \text{d.w.s.}\\
\kol{1}{0}+\kol{0}{1} & = \kol{0}{1}
\ \text{---\ en ni\'{e} }\kol{1}{1}  \ \text{nie!}%
\end{align*}

Trouens, die optelling van die twee vektorbisse werk s\'{o}:%
\[
\kol{a}{b}  +\kol{c}{d}  =\kol{a+d}{a+c} .
\]

Maar kan die voorstelling van omkeerbare line\^{e}re logiese hekke deur
nie-singuliere matrikse nog deurgevoer word met die nuwe vektorvoorstelling
van bisse? Die antwoord is "ja" --- trouens ons is nog beter daaraan toe, want
die \emph{NIE}-hek, byvoorbeeld, wat tevore nie deur 'n matriks voorgestel kon word
nie, kom nou ooreen met ons vroe\"{e}re unit\^{e}re matriks $J,$ wat
$\left\vert 0\right\rangle \ $en $\left\vert 1\right\rangle \ $in mekaar
omsit:%
\[
J \kol{x}{y}  = \left(
\begin{array}
[c]{rr}%
0 & 1\\
1 & 0
\end{array}
\right) \kol{x}{y}  =\kol{y}{x} .
\]
En die beheerde-nie-hek, wat ons tevore deur die matriks $N$ voorgestel
het?\ As ons enigeen van $\left\vert 0\right\rangle $ en $\left\vert
1\right\rangle \ $skryf as, bv.,\ $\left\vert x\right\rangle \ $of\ $\left(
a;b\right)^T  \ $(een van $\left( 1;0\right)^T  \ $of\ $\left(  0;1\right)^T  $),
dan kan ons die hek
\[
\kol{\left\vert x\right\rangle}{\left\vert y\right\rangle} 
\longmapsto  \ckol{\left\vert x\right\rangle}{\left\vert x\right\rangle
+\left\vert y\right\rangle} 
\]
voorstel deur 'n $4\times4$ matriks, aangedui met $BN$ (wat sy eie inverse is):%
\begin{align*}
BN  \kol{\left\vert x\right\rangle}{\left\vert y\right\rangle} 
 & =
\left(
\begin{array}
[c]{rrrr}%
1 & 0 & 0 & 0\\
0 & 1 & 0 & 0\\
1 & 0 & 0 & 1\\
1 & 0 & 1 & 0
\end{array}
\right) \matriks{a}{b}{c}{d} \\
& = \cmatriks{a}{b}{a+d}{a+c} \\
& = \ckol{\ket{x}}{\ket{x}+\ket{y}}.
\end{align*}
Maar nog is het einde niet!
Op die pad na kwantumberekening moet ons die wyse waarop ons die
beheerde-nie-hek 
voorstel nog een stap verder voer. Eers was ons bisse die twee elemente 0 en
1 van 
$\mathbb{Z}_2$ en het ons \emph{BNIE} voorgestel deur die $2\times 2$ omkeerbare
matriks $N$. 
Toe besef ons dat die bisse fisies deur toestande (komponente) van \n  
rekenaar 
gerealiseer moet word. In die kwantumteorie word toestande van \n fisiese
stelsel 
geassosieer met vektore, sodat ons die twee bisse toe sien as vektore $\ken$ en
$\kee$ en \emph{BNIE} voorstel deur die omkeerbare $4\times 4$ matriks $BN$.
Deurgang 
van \n stringetjie van twee bisse deur die beheerde-nie-hek moet fisies
gerealiseer word 
as \n geldige oorgang van een toestand van die fisiese stelsel na \n
volgende toestand. 
Sulke geldige ontwikkelinge van toestande word in die kwantumteorie
wiskundig voorgestel 
deur die toepassing van \emph{unit\^ere} line\^ere transformasies op die
vektor wat 
met die begintoestand geassosieer is. Unit\^er beteken \emph{meer} as net
omkeerbaar. En 
n\'og $N$ n\'og $BN$ is unit\^er! Wat nou?

Om die hek \emph{BNIE} wiskundig deur \n unit\^ere $4\times 4$ matriks voor te
stel, doen die
kwantumteorie aan die hand dat ons die vektorvoorstelling van die vier pare
bisse wat 
ons deur die hek wil jaag, anders benader. Di\'e vier pare, met \n eenvoudige 
skryfwyse, is:
$$\kol{\ken}{\ken} = \ket{00}, \quad \mbox{en soortgelyk $\ket{01}$,
$\ket{10}$ en 
$\ket{11}$.}$$
Sien nou di\'e vier vektore as \n abstrakte ortonormale basis vir \n
vierdimensionele 
vektorruimte, voorlopig oor $\mathbb{Z}_2$, maar later vir kwantumberekening
oor die 
komplekse getalle $\mathbb{C}$. Ons sou die basisvektore, respektiewelik,
netsowel 
$\ket{s}$, $\ket{t}$, $\ket{u}$ en $\ket{v}$ kon noem, waarby ons dan vir
die huidige 
vergeet dat hulle eintlik uit twee komponentvektore saamgestel is. Hierin bemerk 
ons, met \n bietjie goeie wil, reeds \n aanduiding van miskien di\'e
karakteristieke 
eienskap van die kwantumfisika: die \emph{verstrengeling} van toestande, nl.
die feit dat 
die toestand van \n komplekse fisiese stelsel nie noodwendig eenduidig
ooreenkom met 'n 
konfigurasie van toestande van sy samestellende komponente nie.
Verstrengeling is 
miskien die mees presiese illustrasie van die uitdrukking \quotedblbase die geheel is
m\'e\'er as die \quotesinglbase som' van sy dele''.

Wat die beheerde-nie-hek verrig (kyk maar weer na sy waarheidstabel) is die
volgende 
permutasie van die vier basisvektore:
$$\ket{00}\mapsto \ket{00}, \ket{01}\mapsto\ket{01}, \ket{10}\mapsto\ket{11}, 
\ket{11}\mapsto\ket{10}, 
\mbox{of}$$
$$\ket{s}\mapsto\ket{s}, \ket{t}\mapsto\ket{t}, \ket{u}\mapsto\ket{v}, 
\ket{v}\mapsto\ket{u},$$
m.a.w. die laaste twee word net omgeruil. Dieselfde inligting word vervat in
die 
volgende tabel of \textit{matriks}:
\begin{center}
\begin{tabular}{c|cccc}
 & $\ket{s}$ & $\ket{t}$ & $\ket{u}$ & $\ket{v}$ \\
 \hline
 $\ket{s}$ & 1 & 0 & 0 & 0 \\
 $\ket{t}$ & 0 & 1 & 0 & 0 \\
 $\ket{u}$ & 0 & 0 & 0 & 1 \\
 $\ket{v}$ & 0 & 0 & 1 & 0 
 \end{tabular}~,
 \end{center}
waar elke 1 aandui dat die basisvektor(e) in die betrokke ry en kolom deur
die 
transformasie \emph{BNIE} in mekaar oorgevoer word. Dit lewer nou vir ons
uiteindelik \n \emph{unit\^ere} $4\times 4$ matriks \emph{BNIE} wat die beheerde-nie-hek
voorstel as \n 
basistransformasie van \n vierdimensionele vektorruimte: Vir enige vektor
$$\ket{\phi} = \alpha\ket{s} + \beta\ket{t} + \gamma\ket{u} + \delta\ket{v}$$
word sy ko\"ordinaatvektor soos volg getransformeer van die ou na die nuwe
basis:
$$\mbox{\emph{BNIE}}(\alpha;\beta;\gamma;\delta)^T =
\matriks{1 & 0 & 0 & 0}{0 & 1 & 0 & 0}{0 & 0 & 0 & 1}{0 & 0 & 1 & 0}
\matriks{\alpha}{\beta}{\gamma}{\delta} =
\matriks{\alpha}{\beta}{\delta}{\gamma } = (\alpha\;\beta;\delta;\gamma)^T.$$
T.o.v. die ou basis het die ou basisvektore natuurlik die volgende
ko\"ordinaatvektore:
\begin{eqnarray*}
\ket{s} & : (1;0;0;0)^T \\
\ket{t} & : (0;1;0;0)^T \\
\ket{u} & : (0;0;1;0)^T \\
\ket{v} & : (0;0;0;1)^T 
\end{eqnarray*}
Uit die line\^ere algebra weet ons dat ons die matriks \emph{BNIE} ook soos volg
kan kry: Alle 
ko\"ordinaatvektore word t.o.v. die ou basis geneem; dan word die
\emph{beelde} (onder 
die permutasie) van die vier ou basisvektore (in volgorde) se
ko\"ordinaatvektore in die 
vier kolomme van $\mbox{\emph{BNIE}}^{-1} = \mbox{\emph{BNIE}}$ geplaas. \emph{BNIE} se vier
kolomme is dus 
die koördinaatvektore (t.o.v. die ou basis) van $\ket{s}$, $\ket{t}$,
$\ket{v}$, 
$\ket{u}$, in daardie volgorde.

Om verder vertroud te raak met bostaande gedagtegang, skets ons hoe \n
unit\^ere 
$8\times 8$ matriks verkry kan word om die Toffoli- of beheerde-beheerde-nie-hek, \emph{BBNIE}
(\emph{CCNOT} in
Engels), wiskundig voor te stel. Di\'e hek transformeer drietalle bisse,
\[%
\begin{tabular}
[c]{c|c|l}\cline{2-2}%
\begin{tabular}
[c]{c}%
$x\rightarrow$\\
$y\rightarrow$\\
$z\rightarrow$%
\end{tabular}
& $\ T \ $ &
\begin{tabular}
[c]{l}%
$\rightarrow x'$\\
$\rightarrow y'$\\
$\rightarrow z'$%
\end{tabular}
\\\cline{2-2}%
\end{tabular}
\]
volgens die volgende waarheidstabel.
\begin{center}
\begin{tabular}{|ccccccc|}
\hline $x$ & $y$ & $z$ & & $x^{\prime}$ & $y^{\prime}$ & $z^{\prime}$ \\
\hline 0 & 0 & 0 & & 0 & 0 & 0 \\
0 & 0 & 1 && 0 & 0 & 1 \\
0 & 1 & 0 && 0 & 1 & 0 \\
0 & 1 & 1 && 0 & 1 & 1 \\
1 & 0 & 0 && 1 & 0 & 0 \\
1 & 0 & 1 && 1 & 0 & 0 \\
1 & 1 & 0 && 1 & 1 & 1 \\
1 & 1 & 1 && 1 & 1 & 0 \\
\hline
\end{tabular}
\end{center}
Ons merk op dat ons hier twee "beheerdrade" het, met $x$ en $y$, wat
onveranderd deur die hek 
gaan, terwyl die "teiken" $z$, wat deur die derde draad gaan, in $z'=\mbox{nie-$z$}$
verander wanneer 
beide beheerdrade geaktiveer is, $x=1$ en $y=1$; andersins geld $z'=z$.
(Altyd geld $z'=z+xy$. )
Kom ons sien nou hoe lyk die agt drietalle van bisse as ortonormale
basisvektore vir \n 
agtdimensionele vektorruimte en benoem hulle soos volg:
\begin{quote}
\begin{tabbing}
$\ket{p}=\ket{000}$;\quad \=  $\ket{q}=\ket{001}$;\quad \=  $\ket{r}=\ket{010}$;\quad \=  $\ket{s}=\ket{011}$;\\
$\ket{t}=\ket{100}$; \> $\ket{u}=\ket{101}$; \> $\ket{v}=\ket{110}$; \> $\ket{w}=\ket{111}$.
\end{tabbing}
\end{quote}
Dan voer \emph{BBNIE} die volgende permutasie uit:
\begin{quote}
\begin{tabbing}
$\ket{p}\mapsto \ket{p}$; \quad \=  $\ket{q}\mapsto \ket{q}$; \quad \=   $\ket{r}\mapsto \ket{r}$; \quad \=  $\ket{s}\mapsto \ket{s}$;\\
$\ket{t}\mapsto \ket{t}$; \> $\ket{u}\mapsto \ket{u}$; \> $\ket{v}\mapsto \ket{w}$; \> $\ket{w}\mapsto \ket{v};$
\end{tabbing}
\end{quote}
m.a.w. $\ket{v}$ en $\ket{w}$ ruil net om, terwyl al ses die ander
basisvektore
op hulself afbeeld. Dit is duidelik dat die transformasie voorgestel kan word
deur die unit\^ere
$$\mbox{\it BBNIE} = \left(
\begin{array}{cccccccc}
1 & 0 & 0 & 0 & 0 & 0 & 0 & 0 \\
0 & 1 & 0 & 0 & 0 & 0 & 0 & 0 \\
0 & 0 & 1 & 0 & 0 & 0 & 0 & 0 \\
0 & 0 & 0 & 1 & 0 & 0 & 0 & 0 \\
0 & 0 & 0 & 0 & 1 & 0 & 0 & 0 \\
0 & 0 & 0 & 0 & 0 & 1 & 0 & 0 \\
0 & 0 & 0 & 0 & 0 & 0 & 0 & 1 \\
0 & 0 & 0 & 0 & 0 & 0 & 1 & 0 \\
\end{array}
\right).$$
Die bespreking en voorbeelde in hierdie afdeling, ietwat ongewoon in die
konteks van klassieke berekening, het ten doel om aanvoeling en begrip te wek
vir die volgende gedagtegang, wat eers in kwantumberekening ten volle
vergestalt kan word:

\begin{itemize}
\item Die bisse en stringe van bisse, wat die invoere, deurvoere en afvoere
van berekening is, word wiskundig voorgestel deur vektore in 'n gepaste
vektorruimte. Hulle word deur 'n logiese stroombaan gestuur met seri\"{e}le en
parallelle skakeling, van die invoernode af tot by die afvoernode.

\item Die nodes in die stroombaan word wiskundig voorgestel deur omkeerbare
line\^{e}re transformasies, nie-singuliere matrikse, wat op die vektore
toegepas word. In kwantumberekening moet die matrikse unitêr wees en die stroombaan serieel.

\item Enige berekening kan deur 'n gepaste sodanige stroombaan uitgevoer word.

\item Dit is moontlik om fisiese komponente te bou waarvan die toestande die
vektore (bisse en stringe) en hul transformasies realiseer, m.a.w. die
stroombaan vir 'n berekening kan in beginsel deur 'n fisiese stelsel met 'n
eindige aantal standaardkomponente verwerklik word.

\item En dan kom die kwantumteorie met 'n verstommende verrassing:\ 'n enkele
seriële stroombaan kan \textit{gelyktydig }'n groot aantal (eksponensieel in die
aantal bisse) berekenings uitvoer, sonder enige fisiese parallelskakeling van
soveel kopie\"{e} van die stroombaan. So skynbaar vreemd is di\'{e} verskynsel, 
dat sommige kenners in alle erns beweer dat die berekenings in 'n groot aantal
parallelle heelalle moet afspeel!
\end{itemize}

\section{\uppercase{Teoretiese agtergrond}}
\label{TA}

Kwantumteorie is 'n wiskundige model van stelsels in die fisiese w\^ ereld wat
begrippe soos toestande, waarneembares, metings en die dinamika van 'n fisiese
stelsel wiskundig kan voorstel. In kwantumteorie word 'n fisiese stelsel deur 'n Hilbert-ruimte
$\cal H$ voorgestel, terwyl die toestande van die fisiese stelsel deur vektore in
$\cal H$ beskryf word. 'n Waarneembare word deur 'n
selftoegevoegde operator op $\cal H$ voorgestel, terwyl die dinamika van die stelsel deur 'n
kontinue familie $U(t)$ van unitêre operatore beskryf word. Hieronder volg 'n
bondige uiteensetting van die basiese wiskundige konstruksies in die
kwantumteorie. Ons verskaf ook die tipiese fisiese en statistiese vertolkings van die
wiskundige definisies. 
\subsection*{Toestande}
'n {\it Toestand} is 'n volledige beskrywing van 'n fisiese stelsel op 'n
gegewe tydstip. In
kwantummeganika is 'n toestand 'n {\it straal} in die {\it Hilbert-ruimte}
wat die fisiese stelsel verteenwoordig.
Wat is 'n  Hilbert-ruimte?

\begin{enumerate}
\item Dit is 'n vektorruimte oor die komplekse getalle. Vektore
  word aangedui met $|\psi\rangle$ (Dirac se ket-notasie).
\item Dit besit 'n inproduk, $\langle.|.\rangle$, wat elke geordende paar
  $|\phi\rangle,|\psi\rangle$ van vektore na 'n komplekse getal
  $\langle\phi|\psi\rangle$ afbeeld, onderworpe aan die volgende voorwaardes:
\begin{itemize}
\item  $\langle\phi|\phi\rangle \geq 0$, vir alle vektore
  $|\phi\rangle$ en gelykheid geld as en slegs as  $|\phi\rangle$ die
  nul-vektor is;
  
\item $\langle
  \phi|(a|\psi_1\rangle+b|\psi_2\rangle)\rangle=a\langle\phi|\psi_1\rangle+b\langle\phi|\psi_2\rangle$ vir alle komplekse getalle $a$ en $b$;
\item  $\langle\psi|\phi\rangle=\langle\phi|\psi\rangle^*$,
  waar $\langle\phi|\psi\rangle^*$ die komplekse toegevoegde van
  $\langle\phi|\psi\rangle$ aandui.
\end{itemize}
\item Dit is {\it volledig} met betrekking tot die norm $||\psi||=\langle\psi|\psi\rangle^{\frac{1}{2}}$.
\end{enumerate}

Volledigheid is 'n belangrike vereiste in oneindigdimensionele
Hilbert-ruimtes. Ons gaan egter meestal slegs eindigdimensionele
inproduk-ruimtes bespreek, wat in ieder geval volledig is. Dus hoef die
leser nie te bekommerd te wees oor die presiese definisie van volledigheid nie!
 
'n {\it Straal} (Engels: "ray") is 'n ekwivalensieklas van vektore, waar twee vektore as ekwivalent
beskou word as die een van die ander met behulp van vermenigvuldiging met 'n
nie-nul komplekse getal verkry kan word. Dus is
$|\phi_1\rangle$ en $|\phi_2\rangle$ in dieselfde straal as daar 'n nie-nul
komplekse getal $\lambda$ bestaan sodanig dat
\[|\phi_1\rangle=\lambda|\phi_2\rangle.\]
Die fisiese betekenis hiervan is dat $|\phi_1\rangle$ en  $|\phi_2\rangle$
dieselfde kwantumtoestand voorstel. Ons kan altyd 'n verteenwoordiger
$|\phi\rangle$ van 'n straal (van nie-nul vektore) kies om 'n norm een te h\^
e, dit wil s\^ e sodanig dat 
\[\langle\phi|\phi\rangle=1.\]
In hierdie geval s\^ e ons dat $|\phi\rangle$  {\it genormaliseer} is. 

Twee
vektore $|\phi\rangle$ en $|\psi\rangle$ in $\cal H$ word {\it ortogonaal} genoem,
indien $\langle\phi|\psi\rangle=0$. Gestel $\cal H$ is 'n $n$-dimensionele
Hilbert-ruimte. 'n Basis $(|\phi_1\rangle,\dots,|\phi_n\rangle)$ van $\cal H$  word 'n
{\it ortonormale basis} genoem indien al die elemente van die basis
genormaliseer is en indien hulle paarsgewys ortogonaal is, dit wil s\^ e,
indien $\langle\phi_i|\phi_j\rangle=0$ vir alle $i \neq j$.

Ons sal dikwels na 'n nie-nul vektor $|\phi\rangle$ as 'n {\it toestand} verwys. Gegee twee toestande
$|\phi\rangle$ en $|\psi\rangle$, dan kan ons ook die toestand $\alpha|\phi\rangle+\beta|\psi\rangle$ beskou, waar $\alpha$ en $\beta$ enige twee komplekse
getalle (nie beide nul nie) is. Dié lineêre kombinasie stel 'n {\it superposisie} van toestande voor. Slegs die {\it relatiewe}
fase in hierdie superposisie is fisies betekenisvol, dit wil s\^ e slegs die straal voortgebring deur die (nie-nul) superposisie.

\subsection*{Voorbeeld: Kwabisse}
'n Kwantumbis, oftewel 'n {\it kwabis} ({\it qubit} in Engels), is 'n kwantumstelsel met twee basistoestande, waarvan  
die toestand op enige tydstip (met onskadelike dubbelsinnigheid ook "kwabis" genoem) 'n genormaliseerde superposisie van die twee basistoestande is. 
'n Voorbeeld hiervan word gegee deur die kwantumbeskrywing van die
energievlakke van die elektron in 'n waterstofatoom. Die elektron het
,waarskynlikheidsamplitudes' $\alpha$ en $\beta$ om respektiewelik in of die grondtoestand of
'n opgewekte toestand te wees.
Die getalle $\alpha$ and $\beta$ is komplekse getalle.'n Nuttige manier om
hieraan te dink is dat die elektron gedeeltelik in beide energietoestande
bestaan, met waarskynlikheid $|\alpha|^2$ in die grondtoestand en
met waarskynlikheid $|\beta|^2$ in 'n opgewekte toestand. (Hierdie is bloot 'n metafoor en moet
nie te ernstig opgeneem word nie!) Die elektron bestaan beslis en dus moet die
totale waarskynlikheid een wees, wat beteken dat
\[|\alpha|^2 +|\beta|^2=1.\]
Die gesuperponeerde energietoestand van hierdie tweebasistoestand-stelsel word voorgestel deur
\[|\psi\rangle=\alpha|0\rangle+\beta|1\rangle,\]
waar basistoestand $|0\rangle$, respektiewelik $|1\rangle$, aandui dat die elektron in die
grondtoestand, respektiewelik 'n opgewekte toestand, is. Die toestand leef in 'n twee-dimensionele vektorruimte oor die komplekse getalle. Ons kan
aan die twee toestande $|0\rangle$ en $|1\rangle$ as die eenheidsvektore
$(1;0)^T$ en $(0;1)^T$ dink en dus dat $|\psi\rangle$
met $(\alpha;\beta)^T$ ooreenstem. 

Indien
$|\psi_1\rangle=\alpha_1|0\rangle+\beta_1|1\rangle$ en
$|\psi_2\rangle=\alpha_2|0\rangle+\beta_2|1\rangle$ twee kwabisse is, dan word
die inproduk gedefinieer deur:
\[ \langle\psi_1|\psi_2\rangle=\alpha_1^*\alpha_2+ \beta_1^*\beta_2.\]
Dit kan maklik aangetoon word dat 'n kwabis geskryf kan word as
\[|\psi\rangle=\cos\theta|0\rangle+e^{i\phi}\sin \theta|1\rangle.\]
Ons kan aan   $\theta$ en $\phi$ as die sferiese ko\" ordinate van 'n punt op
die eenheidsfeer in die driedimensionele re\" ele ruimte dink. Op hierdie manier kan (genormaliseerde) kwabisse deur die punte
van die eenheidsfeer geparametriseer word.

Dit wil dus voorkom of 'n kwabis 'n oneindige hoeveelheid inligting bevat,
aangesien sy toestand deur twee  {\it kontinue} grade van vryheid voorgestel
word. Dit is egter, ten minste van 'n fisiese oogpunt, {\it glad} nie die geval nie, as
gevolg van die volgende uiters belangrike aspek van kwantumstelsels.

Wanneer 'n kwabis gemeet word, word slegs één van twee resultate verkry; die eiewaarde $\lambda_0$ wat met $\ken$ ooreenkom of die eiewaarde $\lambda_1$ wat met $\kee$ ooreenkom.  Dit werk soos volg: 'n meting van
$|\psi\rangle=\alpha|0\rangle+\beta|1\rangle$ lei tot resultaat $\lambda_0$ met waarskynlikheid
$|\alpha|^2$, en die toestand word $|0\rangle$, of $\lambda_1$ word gemeet met waarskynlikheid
$|\beta|^2$, waarna die toestand $|1\rangle$ is. Vir ons voorbeeld, die
waterstofatoom, meet ons dus die energievlak van die elektron. Met
waarskynlikheid $|\alpha|^2$ sal ons die grondvlak kry, en met waarskynlikheid
$|\beta|^2$ 'n opgewekte vlak. N\' a die meting sal, respektiewelik vir
hierdie twee moontlikhede, die elektron ondubbelsinnig \' of in die grondvlak,
\' of in 'n opgewekte vlak wees. 

Dus lei 'n enkele meting tot slegs een bissie inligting omtrent
$\alpha$ of $\beta$. Dit verg oneindig veel identies voorbereide
kwabismetings om  $|\alpha|$ en $|\beta|$ te bepaal. Dit kom daarop neer dat
oneindig veel metings benodig word om die
relatiewe frekwensies van die uitkomstoestande $|0\rangle$ of
$|1\rangle$ te bepaal n\' a 'n (potensieel) oneindige aantal identiese
voorbereidings van die fisiese stelsel met die twee klassieke makro-(uitkoms-)
toestande $|0\rangle$ en $|1\rangle$ gemaak is. In 'n mate is daar wel 'n klomp verskuilde inligting in 'n
kwabis en soos ons later sal sien, l\^ e die potensi\" ele krag van
kwantumberekening presies hierin, alhoewel dit baie vaardigheid vereis om
toegang tot
hierdie verskuilde inligting te verkry.

\subsection*{Voorbeeld: Multi-kwabisse} 
Beskou twee kwabisse, gerealiseer deur komponente van een fisiese stelsel. Indien  dit deur die twee energietoestande van  elektrone in
twee waterstofatome voorgestel word, dan, klassiek gesproke, is daar vier moontlike toestande, naamlik $00,01,10$ en
$11$ vir die twee elektrone. Die kwantumtoestand assosieer 'n
waarskynlikheidsamplitude met elk van hierdie vier toestande,  sodanig dat die
genormaliseerde toestandvektor vir die twee atome gegee word deur:
\[|\psi\rangle=\alpha_{00}|00\rangle+\alpha_{01}|01\rangle+\alpha_{10}|10\rangle+\alpha_{11}|11\rangle,\]
waar
\[|\alpha_{00}|^2+|\alpha_{01}|^2+|\alpha_{10}|^2+|\alpha_{11}|^2=1.\]
'n Metingsresultaat $x$ met $x \in \{00;01;10;11\}$ kom voor met
waarskynlikheid 
$|\alpha_x|^2$, waarna die stelsel oorgaan na die toestand $|x\rangle$. Indien
ons net
die eerste bis meet, dan kry ons nul met waarskynlikheid
$|\alpha_{00}|^2+|\alpha_{01}|^2$ en die toestand gaan oor na
\[|\psi'\rangle=\frac{\alpha_{00}|00\rangle+\alpha_{01}|01\rangle}{\sqrt{
    |\alpha_{00}|^2+|\alpha_{01}|^2}}.\]
Let op dat $|\psi'\rangle$ genormaliseer word om 'n vektor van lengte (norm) een
te wees.

Beskou nou die algemene geval met $N$  kwabisse. Dink aan 'n fisiese stelsel
$S$ met $N$ komponente, elk waarvan 'n kwabis realiseer, dit wil s\^ e, elk
waarvan by meting nul of een sal lewer. Die kwantumtoestand van die hele $S$ word
gespesifiseer deur $2^N$ komplekse getalle (die
waarskynlikheidsamplitudes), een vir elke rytjie (basisvektor) van $N$ nulle
en ene. Vir $N=200$ is $2^N$ alreeds meer as die
beraamde aantal atome in die heelal! Berekenings met soveel data is byna
onvoorstelbaar, maar, in beginsel, laat die natuur dit toe met slegs 200
atome. Ons kan byvoorbeeld na die energietoestand van die elektrone van 200
waterstofatome kyk. Hierdie geweldige berekeningspotensiaal kan benut word om
berekeningsprosesse te ontwerp, wat uitgevoer word deur gebruik te maak van wat deur die natuur
toegelaat word, in die besonder soos dit in kwantummeganika weerspie\" el word.
\subsection*{Waarneembares en Meting}
'n {\it Waarneembare} is 'n eienskap van 'n fisiese stelsel wat, al is dit bloot in
beginsel,  gemeet
kan word. Die energievlak van die elektron in 'n waterstofatoom is 'n voorbeeld wat ons reeds genoem het. In kwantumteorie word 'n waarneembare as 'n {\it selftoegevoegde
operator} voorgestel. Die wiskunde van hierdie gedagte is soos volg:

'n {\it Operator} $A$ is 'n line\^ ere transformasie op 'n Hilbert-ruimte. Dit
beeld vektore af op vektore en is line\^ er. Dus, as dit 
$|\phi\rangle$ op $|\phi'\rangle$
en  $|\psi\rangle$ op $|\psi'\rangle$ afbeeld, dan word die gesuperponeerde toestand
$a|\phi\rangle+b|\psi\rangle$ deur $A$ op
$a|\phi'\rangle+b|\psi'\rangle$ afgebeeld.

Die {\it toegevoegde} $A^\dagger$ van 'n operator  $A$ word eenduidig bepaal deur
die volgende voorwaarde:
\[\langle A^\dagger \phi|\psi\rangle=\langle\phi|A\psi\rangle.\]
Di\' e gemeenskaplike waarde word aangedui met
\[\langle\phi|A|\psi\rangle.\]
Die toegevoegde sal altyd bestaan as die Hilbert-ruimte eindigdimensioneel
is. Dit kan soos volg bereken word: Indien $A$ voorgestel word deur die  $n \times
n$ matriks $(c_{ij})$, dan word die toegevoegde voorgestel deur die matriks
$(c_{ji}^*)$. Hier gebruik ons die konvensie, soos in algemene gebruik
deur fisici, om die komplekse toegevoegde van 'n komplekse getal $z=x+iy,\;x,y$
re\" ele getalle, met $z^*$, d.w.s. $z^*=x-iy$, aan te dui.

Byvoorbeeld, indien
\[A= \left( \begin{array}{cc}
            0&i\\
            2&1
           \end{array} \right), \] 
dan sal
 \[A^\dagger= \left( \begin{array}{cc}
            0&2\\
            -i&1
           \end{array} \right). \]

'n Operator word {\it selftoegevoeg} genoem indien $A=A^\dagger$. In hierdie
geval word  $\langle\phi|A|\phi\rangle$ die  {\it verwagte waarde} van die
waarneembare $A$ genoem wanneer die kwantumstelsel in die toestand
$|\phi\rangle$ is, op voorwaarde dat die toestand genormaliseer is, dit wil
s\^e  as
$\langle \phi|\phi\rangle=1$. Ons sal later sien dat dit die statistiese
verwagting van 'n eksperimentele meting is.

'n Toestand $|\phi\rangle$ word 'n  {\it eietoestand} van 'n waarneembare $A$
genoem indien daar 'n komplekse getal $\lambda$ bestaan sodanig dat
\[A|\phi\rangle=\lambda|\phi\rangle.\]
In hierdie geval word  $\lambda$ die {\it eiewaarde} van $A$ wat met die eietoestand
$|\phi\rangle$ geassosieer word, genoem. Dit kan aangetoon word (en dit is
baie maklik om te doen) dat die eiewaardes van 'n waarneembare {\it re\" ele}
  getalle is.  Ons sal altyd aanneem dat die eietoestande genormaliseer is.

Twee toestande $|\phi\rangle,|\psi\rangle$ is {\it ortogonaal} indien
$\langle \phi|\psi\rangle=0$. Dit kan aangetoon word dat twee eietoestande van
'n waarneembare  $A$ wat met twee {\it verskillende} eiewaardes van $A$
ooreenstem, altyd ortogonaal is.

Die volgende kom redelik algemeen in kwantumteorie voor: Ons het 'n
waarneembare $A$ op 'n 
$n$-dimensionele Hilbert-ruimte met  $n$ {\it verskillende} eiewaardes $\lambda_1,
\ldots, \lambda_n$ en ooreenstemmende eietoestande $|\phi_1\rangle, \ldots,
|\phi_n\rangle$. 'n Tipiese toestand in hierdie $n$-dimensionele ruimte kan
geskryf word as
\[|\phi\rangle=a_1|\phi_1\rangle+\cdots+a_n|\phi_n\rangle,\]
waar die $a_j$ almal komplekse getalle is. Die $|\phi_j\rangle$ is die  {\it
  eietoestande} wat met die waarneembare $A$ geassosieer is en hulle is dus 'n ortonormale basis vir die Hilbert-ruimte. Die fisiese
  betekenis hiervan is dat die meting van $A$ neerkom op die verval van die
  toestand  $|\phi\rangle$ na een van die eietoestande
  $|\phi_j\rangle$. Die eiewaardes van  $A$ is die moontlike waardes van die
  meting van die waarneembare $A$ terwyl die stelsel in toestand
  $|\phi\rangle$ is. Die waarskynlikheid om die waarde
  $\lambda_j$ te meet word gegee deur $|a_j|^2$. 
Die komplekse koëffisiënte $a_j$ word daarom \textit{waarskynlikheidsamplitudes} genoem. 
Let op dat 
\[A|\phi\rangle=\sum_{j=1}^n\lambda_ja_j|\phi_j\rangle.\]
Voorts word die {\it verwagte uitkoms} van die meting gegee deur die re\" ele getal
\[\langle \phi|A|\phi\rangle =\sum_{j=i}^n\lambda_j|a_j|^2.\]
Dit is die klassieke statistiese verwagting van die uitkoms, gegee uitkomste
$\lambda_j$ met waarskynlikhede $|a_j|^2$.

\subsection*{Voorbeeld}
Beskou die operator $A$ op die kwabisruimte, dit wil s\^e, die ruimte wat deur die vektore 
$|0\rangle,|1\rangle$ onderspan word, waar $A$ vir  $|0\rangle$ op sigself en
vir $|1\rangle$ op $2|1\rangle$ afbeeld. Die eiewaardes is $1$ en $2$ met
ooreenstemmende eietoestande $|0\rangle$ en $|1\rangle$, respektiewelik. Vir 'n
algemene kwabis
\[|\phi\rangle=a|0\rangle+b|1\rangle,\]
sal die meting van die waarneembare $A$ die waarde $1$ lewer met
waarskynlikheid $|a|^2$ en die waarde $2$ sal gemeet word met waarskynlikheid
$|b|^2$. Die verwagte waarde van die meting word gegee deur $|a|^2+2|b|^2$, wat
dieselfde is as $\langle \phi|A|\phi\rangle$.
\subsection*{Toestand-dinamika}
In die voorafgaande voorbeeld, het die meting in die
$|0\rangle,|1\rangle$-basis plaasgevind. Ons kan egter enige ortonormale basis
$|v\rangle,|v^\bot\rangle$ kies en die kwabis daarin meet. Om dit te doen,
skryf die toestand in die nuwe basis as
\[|\phi\rangle=a'|v\rangle+b'|v^\bot\rangle,\]
wat beteken dat die uitkoms-toestand $|v\rangle$ sal wees met 'n
waarskynlikheid $|a'|^2$, of die uitkoms-toestand is
$|v^\bot\rangle$ met 'n waarskynlikheid $|b'|^2$. 

Die line\^ ere afbeelding wat 
$|0\rangle$ op $|v \rangle$ en $|1 \rangle$ op $|v^\bot \rangle$ afbeeld, is
'n voorbeeld van 'n {\it unitêre} operator. Enige line\^ ere operator $U$
wat 'n ortonormale basis op 'n soortgelyke basis afbeeld, word 'n
unitêre operator genoem. 
'n Spesiale geval hiervan is 'n permutasie van 'n ortonormale basis. 
Dit kan aangetoon word dat 'n operator $U$ unit\^
er is as en slegs as
\[U^\dagger=U^{-1},\]
waar $U^{-1}$ die inverse van $U$ is. 
'n Mens kan 'n unitêre operator gebruik om die metingsbasis te
verander. Ons kan ook dié operatore beskou om die verandering van toestande van 'n
kwabis voor te stel.

Dit is 'n fundamentele postulaat in kwantumteorie dat die
tydsdinamika van toestande deur unitêre operatore voorgestel moet word. Dit beteken dat, indien die toestand op tydstip
$t$ gegee word deur $|\psi(t)\rangle$ en die toestand verander oor tyd as
gevolg van eksterne fisiese kragte, dan is daar 'n familie van unitêre
operatore $U(t)$ sodanig dat, vir alle $t$, geld:
\[|\psi(t)\rangle=U(t)|\psi(0)\rangle,\]
waar die familie van unitêre operatore $U(t)$ kontinu van $t$
afhang. Voorts word vereis dat aan die volgende vergelyking voldoen word:
\[U(t)U(s)=U(t+s),\]
vir alle re\" ele $t$ en $s$. Hierdie postulaat hou nou verband met die
vereiste dat die metings-statistiek van 'n kwantumstelsel onafhanklik is van die
ko\" ordinaatstelsels met betrekking waartoe die metings gemaak word.  'n \textit{Kwantumberekening} kan gesien word as 'n
samestelling van 'n eindige aantal unitêre operatore wat op 'n
multi-kwabisruimte inwerk. Ten einde die berekening fisies te implementeer,
moet 'n kontinue familie van unitêre operatore gevind word wat op diskrete
tydstippe die waardes van die eindige aantal unitêre operatore (by die nodes van die berekening se ,stroombaan') aanneem. 'n
Sodanige kontinue familie van unitêre operatore word gevind deur 'n gepaste
Hamiltoniaan (of een of ander geskikte fisiese veld) te bewerkstellig. Dit uitvoering
hiervan bly 'n baie groot uitdaging vir die fisika en die tegnologie.

\subsection*{Saamgestelde stelsels}
'n Enkele kwabis is 'n eenheidsvektor in 'n $2$-dimensionele
Hilbert-ruimte. Veronderstel ons het twee kwabisse wat van mekaar ,onafhanklik' is, maar tog gerealiseer word deur twee komponente van een fisiese kwantumstelsel. Hoe word die saamgestelde
toestande voorgestel? Dit word gedoen deur van die sogenaamde tensorprodukte
van vektorruimtes gebruik te maak. Ten einde tensorprodukte te beskryf, sal
ons voorlopig ons line\^ ere algebra doen sonder om van Dirac se notasie
gebruik te maak. As 
$V$ en $W$ vektorruimtes oor die komplekse getalle is met basisse
$v_1,\ldots,v_n$ en $w_1,\ldots,w_m$, dan is die tensorproduk $V \otimes W$
van $V$ en $W$ 'n $nm$-dimensionele ruimte wat onderspan word deur elemente van
die vorm $v \otimes w$, met $v \in V$ en $w \in W$, die sogenaamde element\^
ere tensore. Die element\^ ere tensore gedra hulself biline\^ er, dit wil s\^ e,
hulle voldoen aan die identiteite
\[\alpha(v \otimes w)=\alpha v \otimes w=v \otimes\alpha w;\]
en
\[u \otimes v+w \otimes v=(u+w) \otimes v;\;\;u \otimes v +u \otimes w= u
\otimes (v+w).\]

'n Basis vir die tensorproduk $V \otimes W$ bestaan uit die $mn$ vektore $v_i \otimes
w_j;\;1 \leq i \leq n, 1 \leq j \leq m$. Dit het die gevolg dat 'n
willekeurige element van $V
\otimes W$ op 'n eenduidige manier geskryf kan word as 
\[\sum_{i,j}\alpha_{ij}v_i \otimes w_j.\]
Die definisie kan op 'n soortgelyke manier na die tensorprodukte van meer as
twee vektorruimtes uitgebrei kan word. 'n Vektor van die vorm $v \otimes w$ word
ook die "tensorproduk" van $v$ en $w$ genoem. Dit is baie belangrik om op te
merk dat nie alle vektore in $V \otimes W$ van hierdie vorm is nie, byvoorbeeld $v_1\otimes w_1+v_2\otimes w_2$ met die $v_i$ en die $w_i$ onafhanklik. 
Baie interessante fisiese verskynsels, waarvan "verstrengeling" ("entanglement") 'n voorbeeld is,
spruit vanuit hierdie eienskap van tensorprodukte.

As $V$ en  $W$ Hilbert-ruimtes is, en die inproduk(te) word aangedui met
$(.\;,\;.)$, dan is die tensorproduk ook 'n Hilbert-ruimte met die ge\" erfde inproduk:
\[(v \otimes w, v' \otimes w')=(v,v')(w,w').\]
Ons keer nou terug na Dirac se notasie en skryf 'n basis van $\mathbb{C}^2
\otimes \mathbb{C}^2$ as
\[|0\rangle \otimes |0\rangle,|0\rangle \otimes |1\rangle, |1\rangle \otimes
|0\rangle, |1\rangle \otimes |1\rangle.\]
Ons skryf dikwels  $|0\rangle \otimes |0\rangle$ as $|0\rangle|0\rangle$ of
selfs 
$|00\rangle$ en soortgelyk vir die ander elemente van die basis.

In die algemeen word 'n  $n$-deeltjie-stelsel deur die tensorproduk van $n$ kopie\" e  van $\mathbb{C}^2$
voorgestel. Let op dat dit 'n  $2^n$-dimensionele Hilbert-ruimte oor die komplekse
getalle is. 'n Tipiese toestand van 'n $n$-kwabisstelsel kan geskryf word as
\[|\psi\rangle=\sum_{x \in \{0;1\}^n}\alpha_x|x\rangle.\]
Hier dui  $\{0;1\}^n$ die versameling van alle bin\^ ere woorde van lengte $n$
aan. Die versameling het $2^n$ elemente. Dit het die implikasie dat die
toestand  van  $n$ onafhanklike kwabisse in 'n kwantumstelsel in 'n $2^n$-dimensionele ruimte
voorgestel word.

\subsection*{Spin}
In die fisika is {\it spin} 'n ,intrinsieke hoekmomentum' wat met mikroskopiese deeltjies geassosieer word. Byvoorbeeld, die {\it spinhoekmomentum} van 'n elektron, langs enige rigting $\hat u$ gemeet, kan slegs een van twee waardes aanneem, naamlik $\frac{1}{2}$ of $-\frac{1}{2}$. Hier is $\hat u$ enige eenheidsvektor in $\mathbb{R}^3$. Ons aanvaar dat fisiese eenhede sodanig gekies is, dat die Planckkonstante dieselfde as $2\pi$ is. Die ooreenstemmende eietoestande word met $|\uparrow_u\rangle$, respektiewelik $|\downarrow_u\rangle$, aangedui. Hierdie toestande word spin-op, respektiewelik spin-af, in die $\hat u$-rigting genoem. Dus, vir 'n gegewe rigting $\hat u$, is daar 'n waarneembare, $\sigma_u$, sodanig dat
$\sigma_u|\uparrow_u\rangle=\frac{1}{2}|\uparrow_u\rangle$ en $\sigma_u|\downarrow_u\rangle=-\frac{1}{2}|\downarrow_u\rangle$.

Elektrone word spin-$\frac{1}{2}$ deeltjies genoem, omdat die grootte van enige elektron se spin 'n half is. Nog voorbeelde van  spin-$\frac{1}{2}$ deeltjies word gegee deur neutrinos, protone en neutrone. Deeltjies met spin besit 'n magnetiese moment, wat eksperimenteel waargeneem kan word as die defleksie van die deeltjies in 'n nie-homogene magnetiese veld (soos by die bekende Stern-Gerlach-eksperiment) of deur die magnetiese velde wat deur die deeltjies self voortgebring word.

Die verband tussen die spin in die $\hat z$-rigting en die $\hat x$-rigting is soos volg:
\begin{equation}
|\uparrow_x\rangle=\frac{1}{\sqrt{2}}(|\uparrow_z\rangle+|\downarrow_z\rangle),
\label{eqn:spinone}
\end{equation}
terwyl die ortogonale toestand gegee word deur
\begin{equation}
|\downarrow_x\rangle=\frac{1}{\sqrt{2}}(|\uparrow_z\rangle-|\downarrow_z\rangle).
\label{eqn:spintwo}
\end{equation}
Vir elk van die twee basiese spin-eietoestande in die $\hat x$-rigting, sal 'n meting langs die $z$-as die resultaat spin-op of spin-af lewer, elk met waarskynlikheid $\frac{1}{2}$.

Beskou nou die samestelling
\begin{equation}
\frac{1}{\sqrt{2}}(|\uparrow_x\rangle+|\downarrow_x\rangle).
\label{eqn:spinthree}
\end{equation}
Hierdie toestand het die eienskap dat, as ons die spin langs die $x$-as meet, dan verkry ons $|\uparrow_x\rangle$ of $|\downarrow_x\rangle$, elk met waarskynlikheid $\frac{1}{2}$. Ons mag nou vra: Wat gebeur as ons die toestand in (\ref{eqn:spinthree}) langs die $z$-as meet?

Indien ons met klassieke bisse te doene sou h\^ e, was die antwoord vanselfsprekend. Die toestand in (\ref{eqn:spinthree}) gaan by meting in een van twee toestande, en in elk met waarskynlikheid $\frac{1}{2}$. Vir {\it elk} van die twee toestande,  is die waarskynlikheid $\frac{1}{2}$ om rondom die positiewe of die negatiewe rigting langs die $z$-as te spin. Dus vind ons spin-op met waarskynlikheid $\frac{1}{2}$ langs die $z$-as.

Dit is egter nie die geval nie! Deur (\ref{eqn:spinone}) en (\ref{eqn:spintwo}) bymekaar te tel, vind ons dat die toestand in (\ref{eqn:spinthree}) niks anders as die toestand \mbox{$|\uparrow_z\rangle$} is nie. Dus sal die meting altyd \mbox{$|\uparrow_z\rangle$} gee. 
Hier sien ons hoe by kwabisse, in teenstelling met klassieke probabilistiese bisse, \textit{waarskynlikheidsamplitudes}, en nie waarskynlikhede nie, op onverwagse maniere bymekaar getel kan word. Ons sien hier, in sy eenvoudigste vorm, die verskynsel ,kwantuminterferensie' in werking.

\subsection*{Verstrengeling}

Beskou twee verskillende spin-$\frac{1}{2}$ deeltjies $A$ en $B$ wat in die
volgende saamgestelde 
toestand is:
\[|\Phi\rangle_{AB}=\frac{1}{\sqrt{2}}(|\uparrow\rangle_A|\downarrow\rangle_B-|\downarrow\rangle_A|\uparrow\rangle_B).\] 
In hierdie uitdrukking stel $|\uparrow\rangle_A,|\downarrow\rangle_A$ en
$|\uparrow\rangle_B, |\downarrow\rangle_B$ die basiese spin-eietoestande van die deeltjies $A$ en $B$ langs die $z$-as
voor. Beskou nou metings van die spin van elk van die deeltjies
langs die $z$-as. Kwantumteorie maak die volgende voorspellings:
\begin{itemize}
\item
Die uitkomste $|\uparrow\rangle_A$ en $|\downarrow\rangle_A$ is ewe
waarskynlik en die metings $|\uparrow\rangle_B$ en $|\downarrow\rangle_B$ is
ook ewe waarskynlik;
\item
die gemete $z$-komponente van die spin van $A$ en $B$ is perfek gekorreleerd,
onafhanklik van die ruimtelike skeiding tussen $A$ en $B$. Byvoorbeeld, as die gemete spin van $A$ spin-op is, dan sal 'n meting van $B$ altyd spin-af wees.
\end{itemize}

In hierdie geval is die toestande van die deeltjies $A$ en $B$ {\it
  verstrengel} ("entangled"), want die saamgestelde toestand $|\Phi\rangle_{AB}$ kan nie as
  $|\phi\rangle_A|\psi\rangle_B$ gefaktoriseer word nie, want andersins, sou
  die spin van $A$ en $B$ definitiewe toestande gehad het wat onafhanklik van
  mekaar is. Daar is dus 'n bepaalde verstrengeling tussen hulle: nie een van
  die twee het 'n toestand van sy eie nie---totdat die spintoestand op enigeen van die twee gemeet word.

Dit is belangrik om daarop te let dat {\it verstrengeling tussen $A$ en $B$ nie op 'n lokale wyse geskep kan word nie}. Die enigste manier wat $A$ met $B$ verstrengel kan word, is deur 'n direkte fisiese interaksie tussen die twee substelsels te bewerkstellig. In teenstelling kan ons die toestand
\begin{equation}
|\uparrow\rangle_A |\uparrow\rangle_B
\label{eqn:spinfour}
\end{equation}
voorberei sonder om $A$ en $B$ in kontak met mekaar te bring. Ons hoef slegs 'n (klassieke!) boodskap na twee eksperimenteerders te stuur met die versoek om elk 'n spin in die $z$-rigting voor te berei. Die enigste manier om 'n verstrengelde toestand soos $|\Phi\rangle_{AB}$ vanuit die toestand (\ref{eqn:spinfour}) te verkry is om 'n kollektiewe unitêre transformasie op (\ref{eqn:spinfour}) toe te pas. Ons kan dit nie verkry deur onafhanklike unit\^ere bewerkings op $A$ en $B$ onderskeidelik toe te pas nie. Ten einde die spintoestande van $A$ en $B$ te verstrengel {\it moet} ons hulle bymekaar bring en toelaat om wedersyds op mekaar in te werk.

Een van die belangrikste dryfvere
  agter kwantumteoretiese inligtingsverwerking  is om verstrengeling
  van kwantumtoestande te bewerkstellig en te benut, want verstrengeling maak die ontwerp van
  merkwaardige algoritmes en  bewysbaar-veilige kriptografiese stelsels
  moontlik. Verstrengeling het ook sy nadele. Inderdaad, verstrengeling van 'n kwantumtoestand met die
  makroskopiese omgewing kan tot 'n onderdrukking van superposisies van
  kwantumtoestande (dekoherensie) lei en dis juis sodanige superposisies wat vir
  die verwagte sukses
  van die kwantumalgoritmes verantwoordelik is. Hierdie probleem word aangepak deur van geskikte
  foutkorreksietegnieke gebruik te maak (\ref{FK}).

\section{\uppercase{'n Model vir kwantumberekening}}
\label{MKB} 

Die basiese berekeningseenhede van kwantumberekening is die multi-kwabisse. 'n
Bin\^ ere woord $q_0 \ldots q_k$ van klassieke berekening word vervang deur die multi-kwabis
\[|q_0 \ldots q_k \rangle =|q_0\rangle \otimes \cdots \otimes |q_k\rangle.\]
Ons sal dikwels 'n natuurlike getal met sy bin\^ ere voorstelling identifiseer.
Ons identifiseer byvoorbeeld $6=1.2^2 +1.2 +0$ met die woord $110$ en op 'n
soortgelyke manier word die toestand $|6\rangle$ deur die multi-kwabis
$|110\rangle$ voorgestel, wat op sy beurt dieselfde is as die tensorproduk
$|1\rangle \otimes |1\rangle \otimes |0\rangle$.

Die vloei van 'n \emph{kwantumberekening} word bepaal deur 'n eindige ry van unit\^
ere transformasies $U(t_i)$ (uit 'n familie $U(t)$ wat kontinu van $t$ afhang) op kwabisse of multi-kwabisse. 'n Unit\^ ere transformasie
op 'n ruimte van multi-kwabisse word 'n {\it kwantumhek} genoem, dit wil s\^ e, 'n
kwantumhek is 'n unitêre transformasie op die tensorproduk van 'n gegewe
maar eindige aantal $2$-dimensionele Hilbert-ruimtes oor die komplekse getalle.

'n Kwantumberekening kan soos volg gevisualiseer word: 'n multi-kwabistoestand
word voorberei wat deur 'n unitêre operator getransformeer word, naamlik
die produk van die eindige ry unitêre transformasies waaraan die toestand
onderwerp word. (Die produk, d.w.s. seriële samestelling, van unitêre transformasies is weer unitêr.) Die
afvoer word verkry deur 'n gepaste waarneembare in die getransformeerde
toestand te meet. 
Die moeilike kuns van \emph{kwantumprogrammering} bestaan daarin dat die regte ry van unitêre transformasies ontwerp moet word, sodat hulle fisies realiseerbaar is, en sodat deur meting van 'n gepaste waarneembare by die afvoernode die korrekte uitkoms van die berekening (eiewaarde of eietoestand van die waarneembare) met sekerheid, of ten minste met 'n aanvaarbaar hoë waarskynlikheid, gevind sal word. Dié waarskynlikheid hoef nie eens so hoog te wees nie, mits daar 'n vinnige algoritme is om die korrektheid van die afvoer te toets (\ref{wyses}).

As 'n illustrasie, beskou die unitêre transformasie op die enkel-kwabisruimte $\mathbb{C}^2$
\begin{eqnarray*}
U : |0\rangle &\mapsto& \frac{1}{\sqrt{2}}(|0\rangle +|1\rangle),\\
U :    |1\rangle &\mapsto& \frac{1}{\sqrt{2}}(-|0\rangle +|1\rangle).
\end{eqnarray*}
Hierdie afbeelding is unitêr, want die vektore 
$U|0\rangle,U|1\rangle$ vorm 'n ortonormale basis van die tweedimensionele Hilbert-ruimte, wat beteken dat $U$ die
standaard 
ortonormale basis op 'n tweede ortonormale basis afbeeld. Laat $S$ die waarneembare wees s\' o dat $S|0\rangle = \frac{1}{2}|0\rangle$ en
$S|1\rangle=-\frac{1}{2}|1\rangle$. Indien ons begin met die kwabis
$|0\rangle$ en daarna $U$ daarop toepas, dan word die
toestand $U|0\rangle=|\psi_0\rangle=\frac{1}{\sqrt{2}}(|0\rangle +|1\rangle)$.
Gestel die waarneembare $S$ word vervolgens in die toestand
$|\psi_0\rangle$ met betrekking tot
die
$|0\rangle,|1\rangle$-basis (die {\it standaard-}basis) gemeet, dan word
die uitkomste
$1/2$ of $-1/2$ verkry, elk met waarskynlikheid $1/2$, en die kwabis is daarna in, respektiewelik en slegs oombliklik, \ket{0} of \ket{1}. Soortgelyk, indien ons met
$|1\rangle$ begin, dan word dit deur $U$ getransformeer na
$|\psi_1\rangle=\frac{1}{\sqrt{2}}(-|0\rangle +|1\rangle)$ en as ons $S$
in hierdie toestand meet, dan verkry ons dieselfde metingstatistiek en eindtoestande.

Let op dat die aksie van $U$ omkeerbaar is. Ons kan die aksie van $U$ omkeer
deur van die unitêre transformasie $U^\dagger$ gebruik te maak. Die meting
is egter 'n onomkeerbare proses. Beide $|0\rangle$ en  $|1\rangle$ lei tot
dieselfde metingstatistiek. 
Hierdie is alreeds 'n interessante kwantumberekening. Dit simuleer 'n
ewekansige muntstuk-gooi-eksperiment. 
In hierdie voorbeeld het die unitêre transformasie 'n {\it aktiewe} rol
gespeel. Dit beskryf die toestandverandering wat deur 'n eksterne opwekking
teweeggebring is. 'n Mens kan ook unitêre transformasies op 'n passiewe
manier benut deur die basis waarin die meting gemaak word, te verander.

Ter illustrasie, laat $S_1= USU^\dagger$. Dit is ook 'n waarneembare met
dieselfde eietoestande $|\psi_j\rangle,\;j=0,1$, en eiewaardes $\pm
1/2$. Indien ons  $S_1$ in die toestand $|0\rangle$ met betrekking tot die
$|\psi_0\rangle,|\psi_1\rangle$-basis meet het ons weereens uitkoms $\pm 1/2$
elk met waarskynlikheid $1/2$ en die kwabis in toestand \ket{\psi_0} of \ket{\psi_1} ná die meting. Dit volg vanuit die identiteite
\begin{eqnarray*}
|0\rangle& =&\frac{1}{\sqrt{2}}(|\psi_0\rangle - |\psi_1\rangle)\\
|1\rangle& =&\frac{1}{\sqrt{2}}(|\psi_0\rangle + |\psi_1\rangle).
\end{eqnarray*}
Dit is 'n voorbeeld van inligtingsverwerking verkry vanuit die meting van 'n
{\it waarneembare} in 'n geskikte raamwerk van basis-eievektore. 

Een van die belangrikste hekke vir kwantumberekening is die sogenaamde {\it
  Hadamard-kwantumhek} $H$. Dit werking van $H$ is soos volg:
\[|0\rangle \mapsto \frac{1}{\sqrt{2}}(|0\rangle +|1\rangle),\]
en
\[|1\rangle \mapsto \frac{1}{\sqrt{2}}(|0\rangle -|1\rangle).\]
Die matriksvoorstelling van $H$ met betrekking tot die standaard-basis
$|0\rangle,|1\rangle$ word gegee deur
\[H=\frac{1}{\sqrt{2}}\left( \begin{array}{cr}
            1&1\\
            1&-1
           \end{array} \right).\]
Die  operator $H$ is unitêr, want dit beeld die standaard ortonormale
$|0\rangle,|1\rangle$-basis op die ortonormale basis
$\frac{1}{\sqrt{2}}(|0\rangle +|1\rangle),\frac{1}{\sqrt{2}}(|0\rangle
-|1\rangle)$ af.

Die twee unitêre transformasies $U,H:\mathbb{C}^2 \rightarrow \mathbb{C}^2$ wat ons sopas bespreek het, is enkel-kwabishekke wat werk op die standaardbasis $|0\rangle,|1\rangle$ van $\mathbb{C}^2$. 'n Mens kan aantoon dat daar ooraftelbaar veel sulke hekke is en dat enigeen van hulle, gesien as unitêre $2 \times 2$ matriks oor $\mathbb{C}$, die vorm
\[e^{ia}\left( \begin{array}{cr}
            \cos b& -i\sin b\\
            -i\sin b& \cos b
           \end{array} \right)\left( \begin{array}{cr}
            \cos c&-\sin c\\
            \sin c& \cos c
           \end{array} \right)\left( \begin{array}{cr}
            e^{-id}&0\\
            0& e^{id}
           \end{array} \right)\]
het, met $a,b,c,d$ enige keuse van vier re\" ele getalle. Vir
\[U= \frac{1}{\sqrt{2}}\left( \begin{array}{cr}
            1&1\\
            -1&1
           \end{array} \right)\]
het ons $a=b=d=0\;$ en $\;c=-\frac{\pi}{4}$; terwyl vir die Hadamardhek $a=d=\frac{\pi}{2}$, $b=0$, en $c=\frac{\pi}{4}$.

Nog 'n belangrike hek is die sogenaamde {\it fasehek} $S$. Die matriksvoorstelling
van $S$ met betrekking tot die standaardbasis word gegee deur
\[S=\left( \begin{array}{cc}
            1&0\\
            0&i
           \end{array} \right),\;a=d=\frac{\pi}{4},\; b=c=0.\]
Dit laat die kwabis $|0\rangle$ onveranderd en beeld $|1\rangle$ af op
$i|1\rangle$. In die algemeen, geld
\[S: \alpha|0\rangle + \beta |1\rangle \mapsto \alpha|0\rangle +i \beta
|1\rangle.\]
Die $T$-{\it hek} is die operator met matriksvoorstelling
\[T=\left( \begin{array}{cc}
            1&0\\
            0&e^{i\pi/4}
           \end{array} \right),\;a=d=\frac{\pi}{8},\;b=c=0, \]
met betrekking tot die standaardbasis.

Die prototipiese multi-kwabiskwantumhek is die {\it beheerde-nie-} of
$\emph{BNIE}$-hek. Hierdie hek het twee invoer-kwabisse; die eerste word die $\it
beheer$-kwabis en die tweede word die {\it teiken}-kwabis genoem. Die werking
van hierdie hek kan soos volg beskryf word: Indien die beheer-kwabis nul is, 
dan bly die teiken-kwabis onveranderd. Indien die beheer-kwabis een is, dan
word die teiken-kwabis omgeruil, met ander woorde:
\[\emph{BNIE}: |00\rangle \mapsto |00\rangle;\;|01\rangle \mapsto
|01\rangle;\;|10\rangle \mapsto |11\rangle;\;|11\rangle \mapsto
|10\rangle.\]
Dit is 'n unitêre transformasie aangesien dit die ortonormale basis
$(|00\rangle,|01\rangle,|10\rangle,|11\rangle)$ van $\mathbb{C}^2 \otimes
\mathbb{C}^2$ permuteer. Die matriks van $\emph{BNIE}$ met betrekkking tot hierdie
basis word (soos in die klassieke geval) gegee deur
\[\emph{BNIE}=\left( \begin{array}{cccc}
            1&0&0&0\\
            0&1&0&0\\
            0&0&0&1\\
            0&0&1&0
           \end{array} \right).\]

Elke unitêre transformasie op die multi-kwabisse is 'n line\^ ere afbeelding
\[U: \bigotimes_{j=1}^n\mathcal{H}_j \rightarrow \bigotimes_{j=1}^n\mathcal{H}_j \]
waar elke $\mathcal{H}_i$ 'n kopie van die toestandruimte $\mathbb{C}^2$ van 'n
enkele kwabis is. Dit is merkwaardig, maar kan bewys word, dat enige sodanige
unitêre transformasie as 'n eindige tensorproduk van unitêre operatore
geskryf kan word, elk waarvan \' of op 'n enkele kwabis inwerk \' of die
$\emph{BNIE}$-hek is. Dit beteken dat die enkel-kwabishekke en die $\emph{BNIE}$-hek as
voortbringers van al die kwantumhekke op multi-kwabisse beskou kan word. Let
op dat die enkel-kwabishekke 'n kontinuum vorm, terwyl, ten einde
kwantumrekenaars te bou, dit nodig is om met 'n diskrete versameling van hekke
te werk. Dit kan aangetoon word dat enige enkel-kwabishek benader kan word tot
en met willekeurige akkuraatheid deur samestellings van 'n vaste eindige versameling hekke. 'n Voorbeeld van so 'n eindige versameling is $\{H;S;T\}$.

\section{\uppercase{Voorbeeld: die Deutsch-Jozsa-algoritme}}
\label{DJ}

In 1992 het David Deutsch en Richard Jozsa die eerste nie-triviale algoritme
vir kwantumberekening gepubliseer. Die doel van di\'e algoritme is om te
bepaal of 'n funksie wat slegs die waardes 0 of 1 kan aanneem op 'n eindige
definisieversameling gebalanseerd \'of identies nul is.  Met
\emph{gebalanseerd} bedoel ons dat die funksie ewe veel keer die waarde 0
aanneem as 1. Die probleem veronderstel dat ons kan aanvaar dat die betrokke 
funksie niks anders as identies nul of gebalanseerd kan wees nie.

Sonder verlies aan algemeenheid neem ons aan dat die definisieversameling van 
die funksies wat ons beskou $\{0;1\}^n$ is, m.a.w. dat dit bestaan uit alle 
woorde van lengte $n$ (daar is natuurlik $2^n$ sulke woorde) oor die
binêre alfabet. In die klassieke geval kan dit gebeur dat ons die
funksiewaarde by die helfte plus een, oftewel $2^{n-1}+1$ van hierdie $2^n$
woorde moet ondersoek alvorens ons 'n beslissing kan maak. 
Kwantumberekening maak dit egter moontlik om---selfs in die ergste
geval---die funksie vir gebalanseerdheid te ondersoek in 'n aantal stappe
wat slegs line\^er groei in $n$ i.p.v. eksponensieel soos in die klassieke
geval. Ons neem aan dat daar 'n orakel\footnote{
In die rekenaarwetenskap is 'n orakel 'n tipe van toorkissie (\emph{black
box}) wat byvoorbeeld 'n bepaalde funksie kan bereken. Die orakel word
toegevoeg aan 'n Turing-masjien en daar word dan ondersoek wat die masjien
kan bereken wanneer dit toegang tot hierdie toorkissie het.
Daar word altyd
presies gedefinieer \emph{wat} die orakel bereken maar glad nie \emph{hoe}
nie. Dié naam is natuurlik ontleen aan die bekende waarsêers van die
klassieke tyd, o.a. die Orakels van Delfi (in Griekeland) en Ammon (in
Egipte).}
beskikbaar is vir die berekening van die funksiewaardes van $f$, sonder
koste aan tyd of ruimte. So 'n orakel kan dan gebruik word om te beslis of $f$ gebalanseerd of identies nul is, maar ten koste daarvan om minstens $2^{n-1}+1$ funksiewaardes te ondersoek. Deutsch en Jozsa wou 'n voorbeeld gee van 'n \emph{egte} kwantumalgoritme wat parallelisme benut en dié algoritme is miskien eerder leersaam as nuttig.

Met \emph{register} hier onder sinspeel ons op 'n \emph{geheueregister} in 'n digitale rekenaar, d.w.s. 'n klein geheue-element waartoe die verwerker vinnige lees- en skryftoegang het. In die geval van 'n kwantumberekening is dit 'n manier om die kwabisse in ons berekening te organiseer in die beskrywing van die algoritme.
Die algoritme om te bepaal of $f:\{0;1\}^n\rightarrow \{0;1\}$ identies 0 is
of gebalanseerd, maak gebruik van 'n tweeregister-kwantumtoestel. Die eerste
register het $n$ kwabisse, almal aanvanklik in die basistoestand $\ket{0}$, en
die tweede register 'n enkele kwabis, ook in $\ket{0}$. Die
eerste stap bestaan daaruit dat die \textit{Walsh-Hadamard-hek} toegepas word op die
eerste register sodat die masjien in die toestand
$$H\ket{0^n} = \frac{1}{2^{n/2}} \sum_{x\in \{0;1\}^n} \ket{x}\ket{0}$$
geplaas word. Nou word die (kwantum-)funksiewaarde bereken deur die kwantumweergawe van die orakel en in die tweede register geplaas, sodat ons die toestand
$$\frac{1}{2^{n/2}} \sum_{x\in \{0;1\}^n} \ket{x}\ket{f(x)}$$
verkry. Dié stap is die kwantumekwivalent van die berekening van $f(x)$ en skryf na die tweede register wanneer $x$ in die eerste register is.
Die beginsel is hier ingeroep dat ons met die kwantum-orakel die funksiewaardes oor 'n gemengde toestand kan bepaal en in die tweede register plaas. Eintlik is hierdie bewerking niks anders as die basiese toepassing van die kwantum-orakel nie. Hoekom? In die kwantumwêreld is berekenings \emph{omkeerbaar} en dus moet ons eers dink hoe 'n klassieke \emph{omkeerbare} orakel vir $n$-bis-invoere sou werk. Dié orakel sou byvoorbeeld $n+1$ bisse  $\left(z_1,z_2,\ldots,z_n,z_{n+1}\right)^T$ as invoer neem en dít transformeer na $\left(z_1,z_2,\ldots z_n,f(z_1,\ldots,z_n)\oplus z_{n+1}\right)^T$. (Hier dui $\oplus$ op optelling modulo 2.) Die kwantum-orakel transformeer eenvoudig $\ket{x}\ket{y}$ na $\ket{x}\ket{f(x)\oplus y}$, wat is wat ons hierbo wou hê!
Vervolgens word die transformasie wat deur die unitêre matriks
$$\kol{1 & 0}{0 & -1}$$
voorgestel word op die tweede register toegepas.
\begin{eqnarray*}
\ket{f(x)}=\ken, f(x)=0 & : & \kol{1 & 0}{0 & -1}\ket{0} = \kol{1 & 0}{0 & -1}\kol{1}{0} =
\kol{1}{0}  =  (-1)^{f(x)}\ket{f(x)};\\
\ket{f(x)}=\kee, f(x)=1 & : & \kol{1 & 0}{0 & -1}\ket{1} = \kol{1 & 0}{0 & -1}\kol{0}{1} =
\kol{0}{-1}  =  (-1)^{f(x)}\ket{f(x)}.
\end{eqnarray*}
Die toestand van die masjien is nou
$$\frac{1}{2^{n/2}} \sum_{x\in \{0;1\}^n} (-1)^{f(x)}\ket{x}\ket{f(x)}.$$ Nou herbereken die masjien $f(x)$ met behulp van die orakel, en plaas
$\ket{f(x)\oplus f(x)}$ in die tweede register. Tegnies gesproke was hierdie nou net weer 'n toepassing van die kwantum-orakel soos hierbo beskryf. Hoewel ons
eintlik nou net die tweede register uitgevee het (\ken gemaak het) is dit
nodig om dit op di\'e manier te doen, omdat die tweede register tydens die
berekening verstrengel is met die eerste register en ons die inbring van die faktor $(-1)^{f(x)}$ nie so direk
apart kan hanteer nie. Die stelsel is nou in die toestand
$$\frac{1}{2^{n/2}} \sum_{x\in \{0;1\}^n} (-1)^{f(x)}\ket{x}\ket{0}$$
en hierdie toestand gaan gebruik word om die afvoer te bepaal deur die
Walsh-Hadamard-hek weer 'n keer op die eerste register toe te pas. Nou, die
Walsh-Hadamard-hek is sy eie inverse, so indien $f(x)=0$ vir alle $x$ dan
verkry ons in die eerste register eenvoudig $\ket{0^n}$. Indien $f$ gebalanseerd is,
dan gebruik ons die feit dat die voorlaaste toestand van die eerste register, 
$$\frac{1}{2^{n/2}} \sum_{x\in \{0;1\}^n} (-1)^{f(x)}\ket{x}$$
loodreg is (in die toestandruimte) op $H\ket{0^n}$ en derhalwe is 
$$H\left(\frac{1}{2^{n/2}} \sum_{x\in \{0;1\}^n} (-1)^{f(x)}\ket{x}\right)
\quad \mbox{loodreg op} \quad H(H\ket{0^n})=\ket{0^n}.$$
Die slotgevolgtrekking is dat indien $f$ gebalanseerd is, sal ons \emph{nooit}
$0^n$ aflees in die eerste register nie, maar indien $f$ identies nul is sal
ons \textit{altyd} $0^n$ aflees in die eerste register aan die einde van die berekening.

Let op dat die Deutsch-Jozsa-algoritme geen probabilistiese aspek het
nie---dit gee deterministies die korrekte antwoord na 'n aantal stappe. Maar
\emph{hoeveel} berekeningstappe dan? Wel, soos vir die klassieke algoritme
neem ons aan dat daar 'n (kwantum-)orakel vir $f$ is. Di\'e algoritme vereis 'n vaste
aantal toepassings van die Walsh-Hadamard-hek, twee evaluerings van $f$ deur die kwantumorakel op
die kwantumstelsel en ten slotte die klassieke kontrole van $n$ bisse in die
eerste register. Die gewenste resultaat word dus verkry in 'n aantal stappe
asimptoties eweredig aan $n$.
Hierdie algoritme toon duidelik aan dat 'n probleem waarvan die
tydkompleksiteit klassiek eksponensieel groei deur middel van kwantumrekene
tot line\^ere tydkompleksiteit gereduseer kan word. Daar bestaan tans maar \'e\'en
ander algoritme---Simon se algoritme---wat eenduidig bewese versnelling van
eksponensi\"ele tyd in die klassieke geval na polinoomtyd in die
kwantumrekene behels.

Navorsers van Innsbruck en die MIT Media Laboratory het in 2003 in die
tydskrif \emph{Nature} bekend gemaak dat hulle daarin geslaag het om di\'e 
algoritme te implementeer met 'n enkele atoom van kalsium in 'n sogenaamde 
ioonval (wat in \ref{subs:tegno} bespreek word).

\section{\uppercase{Eksperimentele uitdagings en ontwerpsprobleme by kwantumberekening}}

Wanneer die probleme by kwantumberekening beskou word---en met die huidige
tegnologie is nog, sover bekend, \emph{geen} nuttige berekening met 'n 
kwantumrekenaar gedoen nie---behoort 'n mens in ag te neem dat
kwantumberekening deels sy ontstaan het in die beskouing van die
probleme wat klassieke berekening in die gesig staar wanneer
stroombane só klein word dat kwantumeffekte nie meer ge\"\i gnoreer kan word
nie. Denkers soos Feynman en Manin het in hierdie probleem 'n uitdaging
gesien en daaruit het die huidige teorie en praktyk van kwantumberekening
gevloei. 
Die mees onmiddellike probleme wanneer daar oor die praktiese
implementering van 'n kwantumrekentoestel gedink word, kan geredelik onder
vier hoofde beskou word:

\subsection{Fisiese realisering van die kwabis}
\label{subs:tegno}

Hierdie probleem---om 'n enkele kwabis of 'n kwabisregister in nagenoeg ge\"\i soleerde toestand daar
te stel sodat dit in berekeninge gebruik kan word, o.a. deur die
Walsh-Hadamard-hekaksie daarop toe te pas---is eintlik opgelos. Daar is 'n
hele aantal maniere om 'n kwantumstelsel met twee klassieke toestande te
isoleer en te manipuleer en die werklike uitdaging is om 'n
kwabis-implementering te kies of te vind wat die ander ingenieursoogmerke
ondersteun. Die versameling tegnieke wat reeds gebruik (kan) word bevat o.a.
die volgende.
\begin{description}
\item[Heteropolimere]
Die eerste kwantumrekentoestel gebaseer op
heteropolimere is reeds in die 1980's ontwerp. In die
heteropolimeer-rekenaar word 'n line\^ere skikking van atome gebruik as
die geheue-selle. Laserpulse van toepaslike frekwensies word gebruik om
bewerkings toe te pas. Weens dipool-interaksies van atome in 'n
heteropolimeer is dit moontlik om individuele atome te adresseer met di\'e
pulse.
\item[Ioonvalle] In 'n ioonvalrekenaar word inligting gekodeer deur die
energietoestande van ione en in die vibrasiemodusse tussen di\'e ione.
Bewerkings op die kwabisse word weereens per laser uitgevoer en elke ioon
word deur 'n afsonderlike laser aangedryf. 'n Enkele berillium-atoom kan
gebruik word om 'n beheerde-nie-hek te implementeer.
\item[KED-holtes] Kwantumelektrodinamiese (KED-) holtes kan gebruik word om
kwabisse, wat fisies deur 'n enkele foton voorgestel word, te manipuleer.
\item[Kernmagnetiese resonansie (KMR)] Spintoestande van atoomkerne in
molekules in 'n vloeistof-ampule in 'n KMR-masjien kan kwabisse voorstel en
ook gemanipuleer word. In hierdie geval word daar eintlik met 'n ensemble van
kwantumrekenaars gewerk, anders as in die implementerings hierbo.
\item[Kwantumkrale] 
Kwantumkrale (Engels: "quantum dots") is baie klein toestelle waarin 'n klein aantal vrye 
elektrone (of moontlik net \'e\'en) vasgevang word. In hierdie toestand word
kwantum-energievlakke aangetref wat gebruik kan word vir die daarstelling
van kwantumbisse.
\item[Kane-rekenaars] 
Bruce Kane, toe nog by die Universiteit van Nieu-Suid-Wallis (UNSW) in Sydney, het
in 1998 voorgestel dat enkele fosfor-atome ingebed word in 'n laag suiwer
silikon, sowat 20nm onder die oppervlak, en dat die spintoestand van die
fosfor-atome gebruik word as kwabisse. Hierdie benadering het twee
hoofvoordele: die dekoherensietyd (\ref{deko})  van die spintoestande van die fosfor-atome
is besonder lank, en die toestande kan met 'n magneetveld gemanipuleer word en
derhalwe in beginsel ge\"\i ntegreer word in bestaande elektroniese toestelle.
Navorsers by UNSW kon teen November 2002 reeds 'n kwabis op hierdie manier
verwerklik. Kane-rekenaars is steeds 'n fokuspunt van navorsing in Australi\"e.
\item[Die Josephson-las] 'n Josephson-las bestaan uit twee supergeleiers wat
geskei word deur 'n nie-supergeleidende laag s\'o dun dat elektrone deur die
isolerende laag kan tonnel (die \emph{Josephson-effek}). Drie
Josephson-lasse kan s\'o opgestel word dat die stelsel twee stabiele
waarneembare kwantumtoestande het wat as kwabis gebruik kan word. 
Bestaande tegnieke vir optiese litografie sou gebruik kon word om 'n
kwantum-stroombaan gebaseer op Josephson-lasse te bou.
\end{description}

\subsection{Beheerde superposisie}

Al die tegnieke vir die daarstelling van 'n kwabis hierbo genoem leen
hulself daartoe om kwantumregisters van verskeie kwabisse op te stel en te
hanteer. In Augustus 2000 het Isaac Chuang en kollegas byvoorbeeld 'n KMR-kwantumrekenaar met vyf
kwabisse gedemonstreer. Omdat KMR 'n gesofistikeerde en goed bemeesterde
tegnologie is, het heelwat van die vooruitgang in kwantumberekening tot
dusver van hierdie tegniek gebruik gemaak. Ioonvalrekenaars (\ref{subs:tegno}) blyk geskik te wees vir die uitvoer van die diskrete
Fourier-transformasie en derhalwe moontlik vir Shor se algoritme (sien onder).
Kwantumkrale en Josephson-lasse het die voordeel dat hulle ge\"\i ntegreer kan word
in gewone stroombane (sonder lasers e.d.m.). In 2003 het NEC en die
Riken-navorsingsinstituut in Japan 'n tweekwabis-toestel met Josephson-lasse
voltooi en 'n beheerde-nie-hek ge\"\i mplementeer. 
Vir 'n Kane-rekenaar is aangetoon hoe 'n stel universele hekke verkry kan
word vir kwantumberekening.

\subsection{Dekoherensie}
\label{deko}

Kwantumdekoherensie kan bes beskryf word as die proses waardeur 'n kwantumstelsel 
wegdryf van sy teoretiese kwantumbeskrywing as gevolg van interaksie met die 
klassieke, makroskopiese w\^ereld. Die mees dramatiese instansie hiervan is met
die proses van meting, wanneer die kwantumtoestand vir \n oomblik gereduseer word tot
\n eietoestand van die operator (waarneembare) wat met die meting geassosieer word. Dit is dan eers na
dekoherensie dat Schrödinger se kat\footnote{
'n Gedagte-eksperiment versin (1935) deur die fisikus Erwin Schrödinger. Die kat word in 'n verseëlde kis geplaas met 'n kwantumtoestel wat met waarskynlikheid 50\% 'n toestand bereik wat veroorsaak dat 'n gif vrygestel word wat tot die dood van die kat lei. Die vraag is wat die toestand binne die kissie kan wees vóór dit oopgemaak word.
} as 't ware regtig in die moeilikheid kan wees. Die dekoherensieproses verklaar waarom
fotone, elektrone ens. in die praktyk duidelike lokalisering het en die kenmerke van
klassieke deeltjies vertoon.

In die tweegleufeksperiment\footnote{
Die tweegleufeksperiment (1805) het oorspronklik bestaan daaruit dat lig deur twee gleuwe diffrakteer word en kringe op 'n skerm vorm. Dié kringe (interferensiepatrone) het donker en ligte dele wat dui op die golf-geaardheid van die lig. Wanneer die eksperiment egter met elektrone uitgevoer word, kan ons ook interferensiepatrone sien. Dít dui op 'n golfkarakter ook van elektrone.
} 
verklaar dekoherensie dat, ten spyte van die interferensiepatroon wat op die plaat
opgemerk kan word, elektrone wel een-vir-een op die plaat inval---soos hulle die 
elektrongeweer verlaat het. Dekoherensie kan vroe\"er op di\'e stelsel geforseer
word deur 'n meting te doen by elk van die gleuwe. In di\'e geval verdwyn die interferensiepatroon 
op die plaat, m.a.w. die kwantumstelsel het a.g.v. die voortydige ingryping \n dramatiese dekoherensie 
ondergaan. Di\'e voorbeeld wys ook hoe dekoherensie be\"\i nvloed kan word deur die gebeure in die
omgewing van die stelsel---in hierdie geval of daar meting by die gleuwe plaasvind al dan nie.

By kwantumberekening wil \n mens uit die aard van die saak dekoherensie vermy---minstens totdat die 
finale meting plaasvind. In die geval van die beheerde-nie-hek met een berilliumatoom in 'n ioonval
word koherensie binne rofweg een \emph{millisekonde} verloor en dit is nie baie moeilik om te besef dat
betekenisvolle berekenings ietwat langer koherensietye mag verg nie.  Omdat foute begin intree in die
berekening lank voordat dekoherensie heeltemal ingetree het, word soms ietwat arbitr\^er gemeen dat
\n enkele toepassing van \n kwantumlogiese hek nie meer as $\frac{1}{10000}$ van die tyd tot dekoherensie behoort te gebruik nie.

Daar is 'n interpretasie van kwantumteorie, die \emph{GRW-model} van Ghirardi, Rimini en Weber, wat postuleer dat 'n kwantumstelsel sporadies spontaan dekohereer, d.w.s. uit 'n verstrengeling van toestande na een van daardie toestande spring, sonder dat enige meting of interaksie met die makroskopiese omgewing plaasgevind het. As dít waar is, dan is die probleem van dekoherensie vir kwantumberekening nog erger as wat ons gemeen het.

\subsection{Kwantumfoutkorreksie}
\label{FK}

Kwantumfoutkorreksie behels metodes om kwantuminligting teen foute wat vanweë
{\it dekoherensie} (\ref{deko}), dit wil s\^ e, verval van interferensie, en ander {\it geruis} ontstaan, te beskerm.
Klassieke foutkorreksie maak van oortolligheid gebruik. Die eenvoudigste
manier is om inligting herhaalde kere te stoor en, as kopieë later van mekaar
verskil, 'n meerderheidstem te neem ten einde die fout te herstel.
Hierdie eenvoudige idee kan nie by kwantuminligting gebruik word nie,
aangesien dit aangetoon kan word dat dit fisies onmoontlik is om 'n kopie van
'n kwabis te maak. Dit volg uit die sogenaamde Kloonverbod-stelling: Daar is
geen kwantumbewerking wat alle  toestande $|\phi\rangle$ na $|\phi\rangle
\otimes|\phi\rangle$ transformeer nie.

Ten spyte hiervan het Peter Shor besef dat, al kan kwantuminligting nie gekopieer
word nie, die inligting van 'n enkele kwabis oor verskeie kwabisse ,versprei'
kan word deur van 'n sogenaamde kwantumfoutkorreksiekode gebruik te
maak. Dus, danksy die ontwerp van Shor, 
as geruis of dekoherensie 'n kwabis bederf, sal die inligting nie verlore
wees nie.

Dit is die geval omdat 'n kode vir kwantumfoutkorreksie sodanig ontwerp is dat 'n sekere
meting, die sogenaamde {\it sindroommeting}, gemaak kan word wat kan vasstel
welke kwabis bedorwe is en boonop op welke van 'n eindige aantal maniere
die fout gebeur het. Dit is 'n verrassende resultaat: Aangesien geruis
willekeurig is, hoe kan die effek van geruis slegs een van 'n eindige aantal
moontlikhede wees? Vir die meeste kodes, kan die effek deur 'n unitêre transformasie beskryf word wat, met betrekking tot 'n geskikte basis, een van die Pauli-matrikse $X,Y,Z$ is. Die rede hiervoor is dat die meting van die sindroom die
projeksie-effek van 'n kwantummeting het. Dus selfs al was die fout vanwe\" e die
geraas willekeurig, kan dit as 'n superposisie van  die Pauli-matrikse
$$\kol{0&1}{1&0},\quad \kol{0&-i}{i&0}\quad \mbox{en} \quad \kol{1&0}{0&-1}$$
tesame met die identiteit $I$ uitgedruk word. Die agterliggende aanname is dat 'n versteuring op 'n kwabis deur 'n unitêre bewerking voorgestel kan word. Die sindroommeting
,dwing' die kwabis om te ,besluit' of 'n sekere ,Pauli-fout' ,gebeur'
het en die sindroom vertel ons welke, sodat dieselfde ,Pauli-fout' gebruik kan
word om die effek van die fout om te keer. Die sindroommeting vertel ons
soveel moontlik omtrent die fout wat plaasgevind het, maar {\it niks} omtrent
die waarde van die kwabis nie---andersins sou die meting enige
kwantumsuperposisie van die kwabis met ander kwabisse vernietig het.

\section{\uppercase{Berekeningstake, -wyses en -tye}}
\label{wyses}

Om enige berekening werklik uit te voer, is 'n fisiese stelsel nodig, soos 'n mens met potlood en papier, of
'n rekenaar. In kwantumberekening, benut die kwantumrekenaar 'n eienskap van die fisiese wêreld wat nooit tevore
in berekening 'n rol gespeel het nie: die bestaan van verstrengelde toestande wat groot hoeveelhede
inligting (simboolstringe) \emph{gelyktydig} (in die onvertakte stroombaan van een kwantumrekenaar) kan voorstel en bewerk, in plaas van afsonderlik. Dit laat mens vermoed dat die belangrikste potensiële nut van kwantumrekenaars mag lê in die uitvoering van berekeningstake wat andersins 'n groot aantal soortgelyke stappe sou verg (óf agtereenvolgens deur dieselfde toestel, óf deur 'n groot aantal soortgelyke toestelle), wat nou gelyktydig deur een toestel uitgevoer kan word---met 'n enorme besparing van tyd.

Om 'n prototipe van só 'n berekeningstaak te kry, dink ons aan die \emph{telefoongids}. Die gids is die eksplisiete beskrywing van 'n bijeksie $t:x\mapsto t(x)$ waar $x$ 'n naam is en $t(x)$ die telefoonnommer(s) van $x$, met die
$x$'e alfabeties georden. Die taak \quotedblbase vir hierdie $x$, vind $t(x)$'' is maklik en kan vinnig uitgevoer word omdat
ons 'n vinnige algoritme het om $x$ in die alfabetiese ordening te vind. Maar dink nou aan die taak wat $t^{-1}$
stel: \quotedblbase 8374608 is 'n $t(x)$, vind $x$''. Omdat die nommers in die gids geen orde vertoon wat vinnige opsporing van 'n nommer moontlik maak nie, is daar geen ander genade as om stelselmatig deur die hele gids te werk totdat 8374608 gevind word, waarna die betrokke $x$ in een stap bereik word. Hier het ons dus 'n bijeksie waarvoor $x\mapsto t(x)$ maklik en vinnig is, terwyl $t(x) \mapsto x$ geweldig tydrowend is om seriegewys te bereken uit die databasis.

Hoe tydrowend? Wel, as die databasis $n$ inskrywings het, dan sal $x$ (vir 'n gegewe $t(x)$) in, gemiddeld gesproke, $\frac{n}{2}$ stappe gevind kan word. Dis sommer kinderspeletjies en kon dalk baie erger gewees het, soos $n^2$ of selfs $2^n$ stappe. Tog is daar selfs in hierdie geval 'n beduidende tydsbesparing moontlik met \textit{Grover se kwantumalgoritme} (1996) vir 'n soektog in 'n ongestruktureerde databasis, wat die antwoord in $\sqrt{n}$ stappe kan kry. Hoe groter $n$ word, hoe groter is die verskil tussen $n$ en $\sqrt{n}$.

'n Ander funksie wat relatief maklik berekenbaar is, is vermenigvuldiging, $\ast: {\mathbb N}\times {\mathbb N} \rightarrow {\mathbb N}$. Terloops, die algoritme om twee natuurlike getalle (of rasionale getalle met 'n eindige desimale voorstelling) te vermenigvuldig wat ons op skool geleer het, is nie eens die mees suinige nie! Maar die omgekeerde van vermenigvuldiging, (priem)faktorontbinding, is erg tydrowend. Alle bekende klassieke algoritmes hiervoor se aantal stappe groei eksponensieel, $a^n$, in die lengte $n$ van die getal se syfervoorstelling.
Maar, soos hier onder bespreek, kan \textit{Shor se kwantumalgoritme} (1994) vir priemfaktorontbinding, die aantal stappe beperk deur 'n polinoomfunksie van $n$, wat volgens konvensie, die taak \emph{haalbaar} ("tractable") maak. Daar is dus voorbeelde van berekeningsprobleme wat klassiek onhanteerbaar word vir groot invoerlengte, maar wat haalbaar is vir kwantumberekening. Ons het egter reeds gewys op die ingenieursprobleme om kwantumalgoritmes fisies te verwesenlik. As voorbeeld kan ons noem dat IBM in 2001 'n eendoelige kwantumrekenaar met 7 kwabisse gebou het, wat met Shor se algoritme $15\mapsto 3\times 5$ korrek kon uitvoer.


Ons beskou nou die plek van kwantumberekening binne die klassieke teorie van
berekening.
'n \emph{Probabilistiese Turing-masjien} (PTM) is 'n masjien waar vir elke
interne toestand van die masjien en toestand van die band nie \'e\'en nie,
maar meerdere (tog, eindig veel) oorgangre\"els gedefinieer is, met 'n gepaardgaande
waarskynlikheidsverdeling oor di\'e oorgangre\"els. Daar word gewoonlik 'n
vaste vertrouensvlak $p>\frac{1}{2}$ gekies (s\^e maar $\frac{2}{3}$) en 
indien 'n spesifieke PTM vir elke spesifieke 
invoer $x$ die afvoer $a(x)$ met waarskynlikheid streng ho\"er as die
vertrouensvlak bereken, dan s\^e ons dat di\'e PTM die funksie $f$, met $f(x)=a(x)$ 
vir alle $x$, bereken.

Dit is duidelik dat 'n mens sonder verlies aan algemeenheid kan
aanneem dat elke waarskynlikheidsverdeling oor di\'e oorgangre\"els slegs
rasionale waardes aanneem. Dan kan ons sien dat enige funksie wat deur 'n
PTM bereken word ook deur 'n tradisionele Turing-masjien (TM) bereken kan
word deur eenvoudig alle moontlike rekenpaaie stelselmatig te deurloop,
totdat bevind word dat vir 'n spesifieke invoer $x$ die nagebootste masjien
die afvoer $f(x)$ gee met waarskynlikheid minstens $p$.  Die vertrouensvlak in 'n
uitkoms kan altyd verhoog word tot 'n willekeurige vlak deur die berekening
te herhaal. Die resulterende vertrouensvlak van die saamgestelde bewerkings 
hang nie af van die grootte van die invoer $x$ nie en derhalwe is 'n PTM-berekening 
niks om op neer te kyk nie!

By probabilistiese algoritmes word onderskei tussen algoritmes van die
\emph{Las Vegas}- en die \emph{Monte Carlo}-tipe. 'n Monte Carlo-algoritme
sluit die berekeningsproses altyd binne 'n redelike tydsverloop af, maar gee
met kleinerige waarskynlikheid 'n verkeerde antwoord. Die algoritmes van die
Las Vegas-tipe gee nooit 'n verkeerde antwoord nie, maar gaan met klein
waarskynlikheid in 'n oneindige rekenproses in. Indien 'n vinnig algoritme
van die Monte Carlo-tipe vir 'n spesifieke probleem bestaan \emph{en}
antwoorde kan maklik nagegaan word vir korrektheid, dan is 'n mens eintlik tevrede, want die
algoritme kan eenvoudig herhaaldelik toegepas word totdat die regte antwoord
verkry word. Die gemiddelde aantal herhalings nodig is min en hang af slegs
van die waarde van die foutwaarskynlikheid en nie van die moeilikheidsgraad
van die oorspronklike probleem nie. Die berekenings van 'n PTM, soos hierbo 
gedefinieer, is---netsoos kwantumberekenings---van die Monte Carlo-tipe.

Dit is nie moeilik om in te sien dat die werking van 'n kwantumrekenaar asook 
di\'e van 'n PTM deur 'n
klassieke Turing-masjien \emph{nageboots} kan word nie, en derhalwe kan 'n
kwantumrekenaar geen funksie bereken wat nie ook in beginsel deur 'n TM
bereken kan word nie. Paul Benioff het in 1982 aangetoon dat daar vir 'n TM
ook 'n kwantum-model bestaan en gevolglik is die klas van funksies wat deur
'n universele TM bereken kan word identies met di\'e wat deur 'n universele
kwantummasjien bereken sal kan word. Die natuurlik vraag is dan waarom mens
sigself dan met kwantumberekening sou besig hou. Die antwoord is eenvoudig:
\emph{spoed}.

In die studie van berekeningskompleksiteit (soos algemeen gebruik in rekenaarwetenskap, wiskunde,
toegepaste wiskunde en in operasionele navorsing) word 'n algoritme vir 'n
funksie $f$ \emph{doeltreffend} genoem indien die aantal primitiewe
berekeningstappe\footnote{In die geval van 'n TM is dít eenvoudig die aantal stappe
wat die masjien uitvoer.} wat uitgevoer moet word om $f(x)$ te bereken vir 'n
bepaalde $x$ begrens kan word vir alle $x$ deur 'n enkele polinoomfunksie van die \textit{lengte}
van die invoer. 
Die klas van \textit{haalbare} probleme, waarvoor daar 'n doeltreffende algoritme bestaan,
word gewoonlik met {\bf P} aangedui.

Die krag van algoritmes wat van die toeval gebruik maak, soos vir 'n
PTM, het aan die wetenskaplike gemeenskap dramaties duidelik geword in 1976
toe die Miller-Rabin-toets vir primaliteit bekend gemaak is. Die
Miller-Rabin-toets het vir die eerste keer 'n \emph{doeltreffende}
probabilistiese algoritme gegee waarmee bepaal kon word (tot op die
onafwendbare klein onsekerheid) of 'n gegewe heelgetal $n$ priem is al dan
nie. In die klassieke berekening was dit op daardie stadium nie moontlik om
primaliteit doeltreffend te bepaal nie: die na\"\i ewe algoritme,
byvoorbeeld, moet $\frac{n}{2}$---wat ni\'e 'n polinoomfunksie van die
lengte van $n$ is nie\footnote{In die binêre alfabet, byvoorbeeld, is die
lengte van $n$ natuurlik $\log_2 n$ en $\frac{n}{2}=2^{\log_2 n -
1}$.}---natuurlike getalle ondersoek vir deelbaarheid in $n$. Gelukkig---of
ongelukkig---het professor Manindra Agarwal van die \emph{Indian Institute
of Technology Kanpur} en twee van sy studente,  Nitin Saxena en Neeraj
Kayal, in Augustus 2002 bekendgemaak dat hulle 'n doeltreffende
\emph{deterministiese} primaliteitstoets ontdek het. Hierdie resultaat is
intussen deur die wiskundige gemeenskap getoets en aanvaar.

Daar bestaan egter steeds probleme waarvoor die beste algoritmes wat bekend
is probabilisties is en werklik die toeval (Engels: \emph{randomness}) uitbuit.
Seker die mees bekende is \emph{Quicksort}, 'n algoritme vir die sorteer
van items in 'n lys, wat 'n daadwerklike versnelling bo sy deterministiese
eweknie bied (hoewel dit met lae waarskynlikheid 'n verkeerde resultaat
gee). Ook in grafiekteorie word met vrug van probabilistiese algoritmes
gebruik gemaak. 

Beskou die probleem van ontbinding van 'n (groot) natuurlike getal in
priemfaktore. Die aanname dat di\'e taak vir 'n hedendaagse rekenaar (asook
op die medium termyn) moeilik oplosbaar is vir 'n groot invoer is,
byvoorbeeld, die grondslag vir die RSA-enkripsieskema waarop 'n rits
kommunikasie-protokolle steun. 
Peter Shor het in 1994 'n kwantumalgoritme vir priemfaktorisering voorgestel.  
Shor se algoritme ontbind (met klein, begrensde, 
waarskynlikheid van 'n fout) die heelgetal wat ingevoer word in
priemfaktore in 'n aantal stappe wat deur 'n polinoomfunksie in die
invoerlengte van bo begrens word. In die klassieke berekening ken ons geen
polinoomtydalgoritme vir priemontbinding nie en baie wetenskaplikes glo dat
daar geen polinoomtydalgoritme vir hierdie probleem bestaan nie. Terloops,
die foutmoontlikheid vir Shor se algoritme is nie baie belangrik nie omdat
die korrektheid van 'n priemontbinding vinnig nagegaan kan word{\ldots} deur
die tentatiewe faktore te vermenigvuldig. S\'o 'n probleem, waarvan 'n
vermeende antwoord doeltreffend nagegaan kan word, behoort tot die klas {\bf
NP} van probleme waarvan 'n oplossing in polinoomtyd \emph{geverifieer} kan word.

Één voorbeeld van só 'n probleem is die \emph{Handelsreisiger-beslissingsprobleem} (HRP). Die doel van dié probleem is om te bepaal, gegee 'n $x$, of daar 'n sirkelroete bestaan wat elkeen van 'n gegewe aantal stede besoek indien die stede en die afstande tussen elkaar bekend is, met totale lengte korter as $x$. Dié probleem is in {\bf NP} omdat 'n mens, gegee 'n sirkelroete-kandidaat, natuurlik maklik kan toets of die roete korter is as $x$ al dan nie! 'n Kwantumalgoritme vir HRP sou 'n sensasionele ontdekking wees, want HRP is bekend {\bf NP}-volledig\footnote{'n Polinoomtydoplossing vir HRP sal ons toelaat om enige NP-probleem in polinoomtyd op te los.}---anders as priemfaktorisering wat \textit{vermoedelik} nie is nie. Kwantum-heuristieke (\textit{nie} kwantumalgoritmes soos hier gespesifiseer) vir HRP is al voorgestel, maar daar bestaan ook klassieke heuristieke vir hierdie probleem.

Shor se algoritme sou 'n instansie kon wees waar 'n
kwantumalgoritme vinniger is as die beste klassieke algoritme vir dieselfde
probleem---maar net indien $\mbox{\bf P} \neq \mbox{\bf NP}$ en indien
priemfaktorisering buite $\mbox{\bf P}$ l\^e. Beide van hierdie
veronderstellings is waarskynlik korrek, maar steeds en sekerlik nog vir
geruime tyd onbewese. Die Deutsch-Jozsa-algoritme en Grover se algoritme vir
die soektog deur 'n ongeordende lys is egter sonder twyfel (asimptoties)
vinniger as hul klassieke eweknie\"e. 
Dit is natuurlik belangrik om uit te wys dat elke
probleem wat in polinoomtyd deur 'n klassieke algoritme opgelos word, wel ook in
polinoomtyd deur 'n kwantumalgoritme opgelos kan word, soos in 1992
aangetoon deur Andr\'e Berthiaume en Gilles Brassard.

Soos wat Richard Feynman in 1982 uitgewys het, is daar kwantumstelsels wat
nie deur klassieke rekenaars nageboots kan word sonder eksponensi\"ele
tydsvertraging nie, maar hierdie probleme is nie van die {\bf NP}-tipe nie.
Die vermoede bestaan onder rekenaarwetenskaplikes dat die klas van probleme 
wat in polinoomtyd deur kwantummasjiene opgelos kan word, {\bf BQP}, groter is as die
ooreenstemmende klas vir PTM'e, {\bf BPP}, maar daar bestaan geen bewys hoegenaamd 
daarvan nie. Trouens, die bevatting
$${\bf P} \subseteq {\bf BPP} \subseteq {\bf BQP} \subseteq {\bf NP}$$
herinner ons dat ${\bf BPP} \neq {\bf BQP}$ sal impliseer ${\bf P} \neq {\bf NP}$,
wat nie net 'n groot mylpaal in die toegepaste wiskundige wetenskappe sal wees nie, 
maar ook vir die outeur(s) van 'n veronderstelde bewys 'n \$1miljoen-prys\footnote{Van 
die Clay-instituut, \url{http://www.claymath.org/}.} sal losslaan.

Kwantumstelsels kan sekere vorme van kriptografie implementeer wat nie klassiek moontlik is nie, byvoorbeeld die uitruil van geheime sleutels oor 'n oop kommunikasiekanaal. Dít laat 'n mens toe om in beginsel boodskappe in absolute geheimhouding uit te ruil, omdat die uitruilstelsel enige poging tot afluistering kan ontdek as gevolg van kwantum-verstrengeling. Hierdie onderwerp, hoewel geweldig interessant en nuttig, is egter slegs raaklynig verwant aan kwantumberekening soos hier bespreek.

\begin{center}
\textbf{\begin{large}Enkele berekeningstake\end{large}}

\begin{tabularx}{0.75\linewidth}{| X | X | X | X |}
\hline \textbf{Taak} & \textbf{Voorbeeld} & \textbf{Klassieke- teenoor kwantumberekening} \\ \hline
Berekening van 'n funksie $f$ (deur middel van 'n formule, byvoorbeeld). & Gegee $x$, bereken $\sin x$. &
Kwantumalgoritmes is nie spesifiek beter of slegter as in klassieke berekening nie.\\ \hline
Probleemoplossing---gegee $f(x)$, vind $x$ of soortgelyk. & Gegee 'n saamgestelde getal, bepaal die priemfaktore daarvan. &
Kwantumberekening skyn daar 'n baie duidelike voordeel in te hou. \\ \hline
Bewysbevestiging---bevestig of die invoer 'n waar stelling is al dan nie & Gegee 'n heelgetal, voer "saamgestel" uit indien die invoer nie priem is nie. &
In sommige gevalle is kwantumberekening eenvoudiger en meer kragtig as klassieke algoritmes.\\ \hline
Kansgetalvoortbringing & Vind 'n kansgetal met 'n gespesifiseerde verdeling, byvoorbeeld 'n heelgetal ewekansig verdeel in 'n interval. & Kwantumalgoritmes kan ware kansgetalle voortbring en gebruik in teenstelling met klassieke algoritmes, wat beperk is tot pseudo-ewekansigheid.\\ \hline
Simulasie van 'n fisiese stelsel & Nabootsing van 'n kwantumstelsel in fisika. & In sekere gevalle, byvoorbeeld vir 'n kwantumstelsel, is die krag van kwantumberekening baie groter.\\ \hline
\end{tabularx}
\end{center}

\section{\uppercase{Slotsom}}

Die chronologie van enkele hoogtepunte in die kwantumberekening lyk só:

\begin{center}
\begin{tabularx}{0.85\linewidth}{l X}
1973& Charles Bennett beskryf 'n \emph{omkeerbare} Turing-masjien. \\
1980& Paul Benioff wys daarop dat die omkeerbaarheid van kwantumevolusie (dit is nou, voor die metingsingryping) impliseer dat rekentoestelle wat op kwantummeganiese beginsels gebaseer is ook omkeerbaar sou moes word. \\
1980& In 'n handboek van Yuri Manin word die idee van 'n kwantumrekentoestel aangeroer. \\
1982& Richard Feynman opper die gedagte dat 'n rekenaar gebaseer op kwantumbeginsels nodig sal wees vir die effektiewe simulasie van kwantumprosesse. \\
1985& David Deutsch beskryf 'n kwantummeganiese Turing-masjien. \\
1992& Die Deutsch-Jozsa-algoritme word beskryf.\\
1994& Peter Shor skep die kwantum-algoritme vir priemfaktorisering. \\
1994--5& Berthiaume, Deutsch en Jozsa en Shor beskryf ontwerpe vir kwantumfoutkorreksie. \\
1996& Lov Grover se kwantumdatabasissoekalgoritme verskyn. \\
2001& By IBM se Almaden-navorsingsentrum word Shor se algoritme vir die eerste keer prakties geïmplementeer---die natuurlike getal 15 word suksesvol gefaktoriseer deur 'n kwantummasjien.
\end{tabularx}
\end{center}

Hoewel kwantumberekening 'n heerlike nuwe werktuig vir berekening is en
moontlik ongekende algoritmiese hulpbronne sal voortbring, stel dit
\emph{nie} 'n nuwe vorm of definisie van \emph{berekening} daar nie. 'n Mens
sou dit kon vergelyk met die invoer van die stoomtrein in die negentiende
eeu: ten spyte van ongekende voordele van spoed, gerief en koste, het dit
geen nuwe roetes oopgemaak nie. (Terloops, op die gebied van vervoer sou
slegs seevaart en ruimtevaart (insluitend lugvaart) kwalifiseer vir di\'e eer!)

Die hoofuitdagings en -geleenthede in kwantumberekening lê in:
\begin{itemize}
\item verdere teoretiese werk oor die grondslae van die kwantumfisika en voortspruitende konseptuele en praktiese probleme van kwantumberekening;
\item ingenieurswerk---die bou van beter werkende kwantumrekentoestelle;
\item algoritmes---die ontwerp van nuwe en interessante algoritmes, veral waar kwantumberekening 'n duidelike voordeel bo klassieke berekening toon; en
\item die oplos van die probleem $\mbox{\bf P} \neq^? \mbox{\bf NP}$, wat die relatiewe krag van kwantumberekening in perspektief sou stel.
\end{itemize} 

\subsection*{\uppercase{Dankbetuiging}}

Ons bedank 'n keurder vir waardevolle opmerkings en wenke, waardeur ons dié artikel beduidend kon verbeter.

\section{\uppercase{Geannoteerde bibliografie}}

\subsection*{Algemene boeke oor kwantumfisika}
\begin{itemize}
\item
Bes, D.R. (2004). \textit{Quantum Mechanics: A Modern and Concise Introductory Course} 
(Springer-Verlag, Berlin \& Heidelberg).
\begin{quote}
'n Bondige (200 bladsye) inleiding tot die kwantummeganika---wiskundige eksak, maar beklemtoon begripmatige helderheid. Een kort hoofstuk (16 bladsye) handel oor aspekte wat kwantumberekening raak.
\end{quote}
\item
Davies, P.C.W. (1987). \textit{Quantum Mechanics} (Routledge and Kegan Paul, London).
\begin{quote}
Slapband-inleiding tot kwantummeganika---nie besonder wiskundig veeleisend nie---deur dié bekende outeur.
\end{quote}
\item
Dirac, P.A.M. (1982). \textit{The Principles of Quantum Mechanics} (Oxford University Press, Oxford; oorspronklik verskyn in 1930).
\begin{quote}
Meesterwerk van Paul Dirac en goeie inleiding tot kwantummeganika. Dié boek, wat die outeur op 28-jarige ouderdom die lig laat sien het, het in sy derde uitgawe in 1947 die moderne \textit{bra}-\textit{ket}-notasie, wat ons ook in hierdie artikel  gebruik, ingevoer.
\end{quote}
\item
Ghirardi, G. (2000). Beyond conventional quantum mechanics, pp. 79-116 in Ellis, J., Amati, D. (eds.) (2000). \emph{Quantum Reflections} (Cambridge University Press, Cambridge).
\begin{quote}
In die navorsingsprogram van Ghirardi, Rimini, Weber en andere word die kwantummeganika uitgebrei met 'n omstrede postulaat oor die spontane stogastiese dekoherensie van stelsels. Die frekwensie van verval van verstrengeling neem toe met die aantal komponente van die stelsel, wat potensieel 'n probleem skep vir 'n groot kwantumrekenaar.
\end{quote}
\item
Schommers, W. (ed.) (1989). \textit{Quantum Theory and Pictures of Reality}, (Springer-Verlag, Berlin).
\begin{quote}
Hoofstuk 1, Evolution of Quantum Theory, gee in 48 bladsye 'n netjiese oorsig van die ontstaan en beslag van die kwantumteorie gedurende die eerste kwart van die twintigste eeu, met aandag aan die verband met klassieke meganika.
\end{quote}
\end{itemize} 

\subsection*{Agtergrond en inleiding tot kwantumberekening}
\begin{itemize}
\item
Bennett C.H. (1973). Logical Reversibility of Computation, \emph{IBM Journal of  Research and Development}. Beskikbaar aanlyn:- \break \url{http://www.aeiveos.com/~bradbury/Authors/Computing/Bennett-CH/} [Toegang op 2003-10-31].
\begin{quote}
Die eerste beskrywing van 'n \textit{omkeerbare} Turing-masjien.
\end{quote}
\item
Bennett, C.H. (1988). Notes on the history of reversible computation, 
\textit{IBM Journal of Research and Development}, 32 (1), 16--23.
\begin{quote}
Bennett het in 1973 as eerste 'n omkeerbare Turing-masjien beskryf. Hy gee 'n oorsig van die termodinamika van inligtingsverwerking tot by die omkeerbare kwantumberekening.
\end{quote}
\item
Deutsch, D. (1998). \emph{The Fabric of Reality} (Penguin Books, London).
\begin{quote}
'n Uiteensetting van die idiosinkratiese filosofie van een van die hooffigure in kwantumberekeningsteorie. Dit gaan oor evolusie, kennis, wiskunde en fisika, met 'n hoofstuk oor kwantumrekenaars, waarin hy die omstrede siening huldig dat 'n kwantumberekening eintlik 'n ontsaglike aantal "klassieke" berekenings in interaktiewe parallelle heelalle (die "multiversum") is.
\end{quote}
\item
Gudder, S. (2003). Quantum computation, \textit{The American Mathematical Monthly}, March 2003, 181--201. As voordruk aanlyn beskikbaar:- \url{http://www.math.du.edu/data/preprints/m0310.pdf} [Toegang 2004-11-29].
\begin{quote}
'n Elementêre inleiding tot die onderwerp---soortgelyk aan dié bydrae, maar bespreek ook Grover se algoritme.
\end{quote}
\item
Johnson, G. (2004). \emph{A Shortcut through Time. The Path to the Quantum Computer} (Vintage, London).
\begin{quote}
'n Begripsmatig toeganklike en helder inleiding wat lekker lees. Die proses van kwantumberekening en die onderliggende fisika word so goed verduidelik as wat moontlik is sonder wiskunde.
\end{quote}
\item
Mermin, D.N. (2003). Copenhagen computation, \textit{Studies in History and Philosophy of Science Part B: Studies in History and Philosophy of Modern Physics}, 34(3), 511--522.
\begin{quote}
Die kwantummeganika wat nodig is om kwantumberekening te verstaan, word uiteengesit ooreenkomstig die sogenaamde "Kopenhagen-interpretasie" van kwantumfisika van Bohr, Heisenberg, Born, Von Neumann en andere. Ons dek dié materiaal in \ref{TA} en \ref{MKB}.
\end{quote}
\item
Mermin, D.N. (2003). From Cbits to Qbits: Teaching computer scientists quantum mechanics, 
\textit{American Journal of Physics}, 71 (1), 23--30. Voordruk aanlyn beskikbaar:- \url{http://za.arxiv.org/abs/quant-ph/0207118} [Toegang op 2004-11-28].
\begin{quote}
'n Soortgelyke aanslag as in "Copenhagen computation" hierbo.
\end{quote}
\item Valiev, K.A. (2005). Quantum computers and quantum computations, \textit{Physics-Uspekhi\textit{} }, 48(1), 1--36. 
Aanlyn beskikbaar:- \url{http://www.turpion.org/php/paper.phtml?journal_id=pu&paper_id=2024} [Toegang 2005-07-10].
\begin{quote}
'n Inleidende oorsig, soortgelyk aan hierdie bydrae, met besondere aandag aan die voorkoming van dekoherensie en aan kwantumfoutkorreksie.
\end{quote} 
\item
\url{http://en.wikipedia.org/wiki/Quantum_computer}
\begin{quote}
Die gratis ensiklopedie wat deur sy gebruikers geskryf word se inskrywing oor kwantumberekening---'n goeie, kort oorsig oor die onderwerp. [Toegang op 2004-11-28].
\end{quote}
\item
\url{http://www.qubit.org/}
\begin{quote}
Webwerf met inleidende materiaal, van die \textit{Centre for Quantum Computation} waar David Deutsch ook werk.\break[Toegang op 2004-11-28].
\end{quote}
\item
\url{http://www.turing.org.uk/turing/} 
\begin{quote}
'n Webwerf oor Turing-masjiene, versorg deur Andrew Hodges, skrywer van die uitstekende biografie\\ 
Hodges, A. (1985). \emph{Alan Turing: The Enigma of Intelligence} (Unwin Paperbacks, London); met ander uitgawes as \emph{Alan Turing: The Enigma}. [Toegang op 2004-11-29].
\end{quote}
\end{itemize} 

\subsection*{Gevorderde oorsigte van kwantumberekening}
\begin{itemize}
\item
Bouwmeester, D., Ekert A., Zeilinger, A. (eds.) (2000). \textit{The Physics of Quantum Information: Quantum Cryptography, Quantum Teleportation, Quantum Computation} (Springer-Verlag, Berlin \& Heidelberg).
\begin{quote}
'n Omvattende gevorderde werk, met besondere aandag aan eksperimentele werk. 
\end{quote}
\item
Gruska, J. (1999). \textit{Quantum Computing} (McGraw-Hill, London).
\begin{quote}
Boek wat deurgaans opdateer word op \url{http://www.fi.muni.cz/usr/gruska/quantum/update.html} 
\break [toegang op 2004-11-29] met byvoegings en korreksies.
\end{quote}

\item Heiss, W.D. (2002). \textit{Fundamentals of quantum information}, Lecture Notes in Physics 68 (Springer-Verlag, Berlin \& Heidelberg).
\begin{quote}
Verrigtinge van die 13de Chris Engelbrecht-Somerskool in Teoretiese Fisika, Devon Valley, Stellenbosch, Januarie 2001.
\end{quote}

\item
Ono, Y.A., Fujikawa, K. (eds.) (2002). \textit{Foundations of Quantum Mechanics in the Light of new Technology (Proceedings of the 7th International Symposium on Foundations of Quantum Mechanics, ISQM-Tokyo'01)} (World Scientific Publishing, Singapore).
\begin{quote}
Binne die hooftema van dié konferensie, Kwantumkoherensie en -Dekoherensie, was tegnologiese vooruitgang se betekenis vir die grondslae en die toepassing van kwantummeganika in die brandpunt. Kwantumberekening en \mbox{-inligting} en die gepaardgaande ingenieursprobleme en -oplossings word uitvoerig belig.
\end{quote}
\item
Kitaev, A.Yu., Shen, A.H., Vyalyi, M.N. (2002). \textit{Classical and Quantum Computation} (American Mathematical Society, Providence, Rhode Island).
\begin{quote}
'n Gevorderde werk wat wyd strek, o.a. met 'n hoofstuk oor kwantumalgoritmes vir probleme in the teorie van Abelse groepe.
\end{quote}
\item
Lo H., Popescu S., Spiller T., (eds.) (1998) \textit{Introduction to Quantum Computation and
Information} (World Scientific Publishing, Singapore).
\begin{quote}
Boek gebaseer op 'n lesingreeks by \textit{Hewlett-Packard Labs} se \textit{Basic Research Institute in the Mathematical Sciences} (BRIMS), in Bristol tussen November 1996 en April 1997. 
Sien veral Popescu, S., Rohrlich, D. se "The Joy of Entanglement" en bydraes deur Bennett, Chuang, Jozsa e.a.
\end{quote}
\item
Macchiavello, C., Palma, G.M., Zeilinger, A. (eds.) (2001).
\textit{Quantum Computation and Quantum Information Theory} (World Scientific Publishing, Singapore).
\begin{quote}
Herdrukke van meer as 50 belangrike artikels word in één handige bron byeengebring. Al die aspekte wat ons bespreek, word aangeraak.
\end{quote}
\item
Nielsen, M.A., Chuang I.L. (2000). \emph{Quantum Computation and Quantum Information} (Cambridge University Press, Cambridge).
\begin{quote}
'n Omvattende gevorderde standaardwerk, met 'n insiggewende inleiding.
\end{quote}
\item
Steane, A.M. (1998). Quantum computing, \textit{Reports on Progress in Physics}, 61, 117--173. 
\break Aanlyn beskikbaar:- 
\url{http://xxx.lanl.gov/abs/quant-ph/9708022} [Toegang op 2004-11-28].
\begin{quote}
Hierdie oorsigartikel dek nie alleen kwantumberekening nie, maar ook ander aspekte van die fisika van kwantuminligting, soos kwantumverplasing.
\end{quote}

\item 
\url{http://www.vcpc.univie.ac.at/~ian/hotlist/qc/research.shtml}
\begin{quote}
'n Omvattende lys van navorsing oor kwantumberekening en verwante sake, ingedeel volgens lande [Toegang op 2005-06-14].
\end{quote}
\end{itemize}

\subsection*{Aanvange, uitvoerbaarheid en voordele van kwantumberekening}
\begin{itemize}
\item
Brown, L.M. (ed.) (2000). \textit{Selected Papers of Richard Feynman with Commentary} (World Scientific Publishing, Singapore).
\begin{quote}
Deel VII van die boek---"Computer Theory"---\begin{scriptsize}\begin{footnotesize}\end{footnotesize}\end{scriptsize}bevat Feynman se artikels oor kwantumberekening.
\end{quote}
\item
Deutsch, D. (1985). Quantum theory, the Church-Turing Principle, and the universal quantum computer, \emph{Proceedings of the Royal Society of London}, A400, 96--117. Aanlyn beskikbaar:- \url{http://www.qubit.org/oldsite/resource/deutsch85.pdf}\break [Toegang op 2004-11-28].
\begin{quote}
David Deutsch beskryf hierin vir die eerste maal 'n kwantummeganiese Turing-masjien, ja---meer omstrede---'n \textit{universele} kwantumrekenaar.
\end{quote} 
\item
Deutsch, D., Hayden, P. (1999). Information flow in entangled quantum systems, \textit{Proceedings of the  Royal Society of London}, A456, 1759--1774.
\begin{quote}
Redeneer dat inligting in kwantumstelsels gelokaliseerd is, Bell se Stelling ten spyt.
\end{quote}
\item
Di Vincenzo, D.P. (1995). Two-bit gates are universal for quantum computation. \textit{Physical Review A}, 51 (2), 1015--1022.\break Aanlyn beskikbaar:- \url{http://citeseer.ist.psu.edu/divincenzo95twobit.html} [Toegang op 2004-11-05].
\begin{quote}
Daar word bewys dat kwantumhekke wat op slegs \textit{twee} kwabisse inwerk voldoende is om 'n algemene kwantumstroombaan op te bou.
\end{quote}
\item
Feynman, R. (1982). Simulating physics with computers. \textit{International Journal of Theoretical Physics}, 21 (6\&7),
467--488. Aanlyn beskikbaar:- \url{http://citeseer.ist.psu.edu/feynman82simulating.html} [Toegang op 2004-11-05].
\begin{quote}
Een van Feynman se bekende baanbreker-artikels oor die fisika van berekening.
\end{quote}
\item
Feynman, R.P. (1986). Quantum mechanical computers, \emph{Foundations of Physics}, 16, 507--531.
\begin{quote}
'n Meer gedetailleerde uiteensetting van sy idees.
\end{quote}
\item
Feynman, R.P. (2000). \textit{The Feynman Lectures on Computation} (Perseus Publishing, New York; oorspronklik in 1996, Addison Wesley, Reading, Massachusetts).
\begin{quote}
Hy bespreek klassieke berekening en rekenaars, kodering en inligtingsteorie, omkeerbaarheid en die termodinamika van berekening, en kwantumberekening en -rekenaars.
\end{quote}
\item
Jozsa, R. and Linden, N. (2003). On the role of entanglement in quantum computational speed-up, 
\textit{Proceedings of the Royal Society of London}, A459 (2036), 2011--2032.
\begin{quote}
Verduidelik die wesenlike rol van verstrengeling in parallelle prosessering en gevolglike tydsbesparing.
\end{quote}
\item
Lloyd, S. (23 Augustus 1996). Universal quantum simulators,
\emph{Science}, 273(5278), 1073--1078.
\begin{quote}
Lloyd bevestig die teoretiese steekhoudendheid van Feynman (1982) se simulering van kwantumstelsels deur kwantumrekenaars.
\end{quote}
\item
Potgieter, P.H. (2004). Die voorgeskiedenis van kwantumberekening, \textit{Suid-Afrikaanse Tydskrif vir Natuurwetenskap en Tegnologie}, 23 (1/2), 2--6.
\begin{quote}
Kort maar omvattende oorsig oor die ontstaan van die idees in kwantumberekening, met die klem op verwikkelings in die Sowjet-Unie.
\end{quote}
\item
\url{http://kh.bu.edu/qcl/} 
\begin{quote}
Argief van beduidende artikels oor kwantumberekening, instandgehou deur Tom Toffoli en Zac Walton. Baie artikels afgetas uit die oorspronklike bronne.
\end{quote}
\end{itemize}

\subsection*{Kwantumalgoritmes}
\begin{itemize}
\item
Deutsch, D., Jozsa, R. (1992). Rapid solutions of problems by quantum computation,  \textit{Proceedings of the Royal Society of London}, A439, 553--558.
\begin{quote}
Die Deutsch-Jozsa-algoritme wat in \ref{DJ} bespreek word.
\end{quote}
\item 
Ekert, A. en Jozsa, R. (1996). Shor's quantum algorithm for factoring numbers, \textit{Reviews of Modern Physics}, 68, 733--753.
\begin{quote}
'n Redelik toeganklike uiteensetting van Shor se faktoriseringsalgoritme.
\end{quote}
\item
Grover, L. (1996). A fast quantum mechanical algorithm for database search,
\emph{Proceedings of the 28th Annual ACM Symposium on Theory of Computing}, 212--219. Aanlyn beskikbaar:- \url{http://za.arxiv.org/abs/quant-ph/9605043} [Toegang op 2004-11-28].
\begin{quote}
Grover se Algoritme om 'n item in 'n ongestruktureerde databasis met $n$ items in $\sqrt{n}$ stappe te vind. ACM is die \emph{Association for Computing Machinery}.
\end{quote}
\item 
Shor, P.W. (1997).
Polynomial-time algorithms for prime factorization and discrete logarithms on a quantum computer,
\emph{SIAM Journal on Computing}, 26 (5), 1484--1509.
\begin{quote}
SIAM is die \emph{Society for Industrial and Applied Mathematics}. Dis 'n verbeterde en uitgebreide weergawe van die artikel uit 1994 waarin Shor se Algoritme oorspronklik beskryf is.
\end{quote}
\item
Vandersypen, L.M.K. et al. (2000). Implementation of a three-quantum-bit search algorithm,
\emph{Applied Physics Letters}, 76 (5), 646--648.
\begin{quote}
Klein verwerkliking van Grover se soekalgoritme vir ongestruktureerde databasisse met 'n molekulêre kwantumrekenaar met 3 kwabisse.
\end{quote}
\item
Vandersypen, L.M.K. et al. (20/27 Desember 2001). Experimental realization of Shor's quantum factoring algorithm using nuclear magnetic resonance,
\emph{Nature}, 414 (6866), 883--887.
\begin{quote}
Verwerkliking van die priemfaktorisering $15=3\times 5$ met Shor se Algoritme op 'n eendoelige kwantumrekenaar met 7 kwabisse.
\end{quote}
\end{itemize}

\subsection*{Kwantumfoutkorreksie}
\begin{itemize}
\item
Berthiaume, A., Deutsch, D., Jozsa, R. (1994). The stabilisation of quantum computations,
\emph{Proceedings of the Workshop on Physics and Computation, PhysComp94}, (IEEE Computer Society, Los Alamitos, California), 60--62.
\begin{quote}
Een van die eerste publikasies nadat die teoretiese insig deurgebreek het dat foutkorreksie vir kwantumberekening moontlik is.
\end{quote}
\item
Knill, E. et al. (2001). Bench-marking quantum computers: The five-qubit error correcting code,
\emph{Physical Review Letters}, 86(25), 5811--5814.
\begin{quote}
Manny Knill en andere het dié eksperiment met kernmagnetiese resonansie by Los Alamos uitgevoer. Binne een molekuul is een kwabis verstrengel met vier ander. Met fynontwerpte opeenvolgende elektromagnetiese pulse is 'n doelbewuste fout ingevoer, opgespoor en gekorrigeer.
\end{quote}
\item
Shor, P.W. (1995). Scheme for reducing decoherence in quantum computer memory,
\emph{Physical Review A}, 52 (4), 2493--2496.
\begin{quote}
Vroeë werk oor foutkorreksie in kwantumberekening.
\end{quote}
\item
Steane, A.M. (1996). Error correcting codes in quantum theory,
\emph{Physical Review Letters}, 77(5), 793--797.
\begin{quote}
Die outeur bring die basiese kwantumteorie in verband met die klassieke
inligtingsteorie.
\end{quote}
\end{itemize} 

\end{document}